\documentclass[fleqn,usenatbib]{mnras}

\usepackage{newtxtext,newtxmath}

\usepackage[T1]{fontenc}

\DeclareRobustCommand{\VAN}[3]{#2}
\let\VANthebibliography\thebibliography
\def\thebibliography{\DeclareRobustCommand{\VAN}[3]{##3}\VANthebibliography}

\usepackage{graphicx}	
\usepackage{amsmath}	

\title{Exploring the convective core of the high-amplitude $\delta$ Scuti star TIC 120857354 with asteroseismology}

\author[Chen, Zhang, \& Li]{
Xinghao Chen$^{1,2,3}$\thanks{E-mail:chenxinghao@ynao.ac.cn}
Xiaobin Zhang$^{4}$
and Yan Li$^{1,2,5,6}$
\\
$^{1}$Yunnan Observatories, Chinese Academy of Sciences, P.O. Box 110, Kunming 650216, China\\
$^{2}$Key Laboratory for Structure and Evolution of Celestial Objects, Chinese Academy of Sciences, P.O. Box 110, Kunming 650216, China\\
$^{3}$International Centre of Supernovae, Yunnan Key Laboratory, Kunming 650216, P. R. China\\
$^{4}$Key Laboratory of Optical Astronomy, National Astronomical Observatories, Chinese Academy of Sciences, Beijing, 100012, China\\
$^{5}$University of Chinese Academy of Sciences, Beijing 100049, China\\
$^{6}$Center for Astronomical Mega-Science, Chinese Academy of Sciences, 20A Datun Road, Chaoyang District, Beijing, 100012, China     
}

\pubyear{2015}

\begin{document}
\label{firstpage}
\pagerange{\pageref{firstpage}--\pageref{lastpage}}
\maketitle

\begin{abstract}
Based on 2-minute cadence TESS data, 20 confident independent frequencies were identified for the star TIC 120857354. The Kolmogorov-Smirnov test reveals a rotational splitting of 2.40 $\mu$Hz and a uniform frequency spacing of 74.6 $\mu$Hz. Subsequently, five sets of rotational splittings were discerned, including a quintuplet and four pairs of doublets, aligning with the characteristics of p-mode rotational splitting. Based on the sets of rotational splittings and the uniform frequency spacing, we finally identified 4 radial modes, 6 dipole modes, and 10 quadrupole modes. Furthermore, we found that the frequency separations within the $\ell$ = 2 sequences show a decreasing trend towards lower-order modes, analogous to the $\ell$ = 0 sequences. A grid of theoretical models were computed to match the identified frequencies, revealing that TIC 120857354 is a main-sequence star with $M$ = 1.54 $\pm$ 0.04 $M_{\odot}$, $Z$ = 0.015 $\pm$ 0.003, $T_{\rm eff}$ = 7441 $\pm$ 370 K, $\log g$ = 4.27 $\pm$ 0.01, $R$ = 1.52 $\pm$ 0.01 $R_{\odot}$, $L$ = 6.33 $\pm$ 1.53 $L_{\odot}$, age = 0.53 $\pm$ 0.07 Gyr, and $X_c/X_0$ = 0.84 $\pm$ 0.05. In-depth analyses suggest that $\ell$ = 2 may be p-dominated mixed modes with pronounced g-mode characteristics, enabling us to probe deeper into interiors of the star and determine the relative size of the convective core to be $R_c/R$ = 0.092 $\pm$ 0.002.
\end{abstract}




\section{Introduction}
Thanks to the space missions CoRoT (Baglin et al. 2006), Kepler (Borucki et al. 2010), and TESS (Ricker et al. 2015), thousands of high-precision and long-duration brightness measurements have been acquired. Such data enable us to probe the physics of stellar interiors via asteroseismology, further advancing our understanding of stellar internal structure and evolution, such as convection and rotation (Aerts 2021).

The $\delta$ Scuti variables are a category of  A- and F-type stars that locate in the overlap region between the main sequence band and the lower extension of the Cepheid instability strip. Their pulsation periods range from 0.02 to 0.25 days (Breger 2000), primarily driven by the $\kappa$ mechanism (Baker $\&$ Kippenhahn 1962, 1965; Zhevakin 1963; Li $\&$ Stix 1994) operating within the second partial ionization zone of helium (Chevalier 1971; Dupret et al. 2004; Grigahc{\`e}ne et al. 2005). In addition, the  influence of turbulent pressure in exciting oscillations has been investigated (Houdek 2000, 2008; Antoci et al. 2014, 2019; Xiong et al. 2016). Moreover, the edge-bump mechanism has also been observed in some chemically peculiar stars (Stellingwerf 1979; Murphy et al. 2020). Both radial and non-radial modes have been observed, making $\delta$ Scuti stars highly significant and promising candidates for asteroseismology.

Mode identification of $\delta$ Scuti stars is a long-standing challenge, since their oscillations do not fall in the asymptotic regime due to the low radial order. Over the past four decades, four primary methods of mode identification have been applied to $\delta$ Scuti stars. The first two methods are the spectroscopic method (Mantegazza 2000) and the multicolor photometric method (Watson 1988), respectively. Give that most of modes exhibit low amplitude, the application of the two methods is significantly constrained by the requirement for high temporal resolutions and short exposure times. Meanwhile, due to the intricate frequency spectra, only a limited number of modes can be identified using these two methods, such as for FG Vir (Daszynska-Daszkiewicz et al. 2005; Zima et al. 2006) and 4 Cvn (Castanheira et al. 2008; Schmid et al. 2014; Breger et al. 2017). The third method is based on the principle of rotational splitting. The rotational splitting gives rise to multiplet structures in the frequency spectrum, such as $\ell$ = 1 triplets and $\ell$ = 2 quintuplets. Frequencies can be identified when they are detected in rotational multiplets, such as KIC 9700322 (Breger et al. 2011), KIC 11145123 (Kurtz et al. 2014), HD 50844 (Chen et al., 2016), CoRoT 102749568 (Chen et al., 2017), and EE Cam (Chen \& Li, 2017). The perturbative methods for analyzing rotational splittings are only viable for slow rotation (Ouazzani et al. 2010). The fourth method primarily employs the regularities in frequency spacings to identify the pulsation modes of $\delta$ Scuti stars. Papar$\acute{\rm o}$ et al. (2016a, 2016b) reported numerous distinct series of regular frequency spacings in 77 $\delta$ Scuti stars. Bedding et al.(2020) identified regular sequences of $\ell$ = 0 and 1 for high-frequency oscillation modes in 60 intermediate-mass main-sequence stars via frequency $\acute{\rm e}$chelle diagrams. Using this method, numerous frequency sequences were subsequently identified in other young $\delta$ Scuti stars (Murphy et al. 2021, 2022, 2023). Nevertheless, those frequency sequences mainly contain $\ell$ = 0 and $\ell$ = 1 modes, while higher-degree sequences are often absent, such as $\ell$ = 2. Furthermore, the scaling relationship between the low-order frequency separation and the stellar mean density (Su\'arez et al. 2014; Garc\'ia Hern\'andez et al. 2015, 2017), the amplitude modulation (Bowman \& Kurtz 2014; Barcel\'o Forteza et al. 2015; Bowman et al. 2016), and the empiric relationship among surface gravity, the effective temperature, and the frequency of the maximum oscillation power $\nu_{\rm max}$ (Barcel\'o Forteza et al. 2018, 2020) have also been investigated. Recently, Bedding et al. (2023) and Murphy et al (2024) demonstrated that  $\nu_{\rm max}$ does not exhibit a strong correlation with stellar temperature. Li et al. (2024) showed a discernible relation between frequency and the temperature in a narrow colour range. Consequently, the usefulness of a $\nu_{\rm max}$-$T_{\rm eff}$ relation is not agreed upon. These patterns and mode identification strategies greatly enrich our understanding of $\delta$ Scuti stars and provide us new insights for exploring stellar interiors and underlying physical processes.

For the high-amplitude $\delta$ Scuti stars (HADS), characterized by amplitudes that approximate or exceed 0.1 magnitude (Petersen 1989; Petersen \& Christensen-Dalsgaard 1996), mode identifications become more clear and straightforward because these stars typically display oscillations in the radial fundamental mode and/or low overtone radial modes (McNamara 2000), such as AI Vel (Walraven et al. 1992), V974 Oph (Poretti 2003), GSC 00144-03031 (Poretti et al. 2005), VX Hydrae (Xue et al. 2018), and GSC 4552-1498 (Sun et al. 2021). Moreover, Walraven et al. (1992) and Poretti (2003) reported that nonradial modes can be present in HADS stars. Thanks to various space missions, large number of low-amplitude frequencies have been detected in HADS, such as V2367 Cyg (Balona et al. 2012), CoRoT 101155310 (Poretti et al. 2011), TIC 308396022 (Yang et al. 2021), and TIC 65138566 and TIC 139729335 (Chen et al. 2024).

TIC 120857354 is a high-amplitude $\delta$ Scuti star characterized by a predominant frequency of 20.49 c/d (Clang et al. 2013; Barac et al. 2022). The dominant frequency suggests that TIC 120857354 is a young main sequence star. Besides, HADS generally rotate slowly with $\upsilon\sin$i $\le$ 30 km/s (Breger 2000). The slow rotation enable the investigation of rotational multiplets according to works of Saio (1981), Dziembowski \& Goode (1992), and Aerts et al. (2010). In this study, we try to identify its frequency spectrum through rotational splitting and regular frequency spacing, aiming to gain a deeper understanding of the intricate oscillation behavior of $\delta$ Scuti stars and to uncover their evolutionary history and internal structure. Section 2 details the observation data and the process of frequency extraction. Section 3 devels into the specifics of mode idnetification. Following this, Section 4 delineates the results of the asteroseismic modeling. Finally, we conclude and discuss our findings in Section 5.

\section{Observations and frequency extraction}

TIC 120857354 was observed by the TESS satellite in the 2-minute cadence mode during Sectors 36 and 37, from 2021 March 7 to 2021 April 28. The photometric data were downloaded from the Mikulski Archive for Space Telescopes (MAST, https://mast.stsci.edu/portal/Mashup/Clients/Mast/Portal.html). In the dataset, we used the times in Barycentric Julian Date (BJD) format and used the "PDCSAP$\_$FLUX" values that processed by the Pre-search Data Conditioning Pipeline (Jenkins et al., 2016) to remove instrumental trends. Outliers were removed from the light curves and the fluxes were normalized following the methodology of Slawson et al. (2011). Figure 1 displays the resultant light curves, revealing that  the star is variable with a peak-to-peak amplitude of  about 0.12 magnitudes.

We carried out a multiple-frequency analysis of the photometric data using the Period04 program (Lenz \& Breger, 2005). Throughout this process, we employed an empirical threshold of S/N = 4 to deem a frequency significant (Breger et al., 1993). Significant peaks were sequentially extracted using the pre-whitening method, and 51 frequencies exceeding the S/N threshold were obtained. The amplitude spectra are detailed in Figure 2, with the extracted frequencies listed in Table 1. Noise was assessed within the surrounding 2 day$^{-1}$ frequency range, and uncertainties in both frequencies and amplitudes were determined through Monte Carlo simulations following the approach of Fu et al. (2013).

Subsequently, following P{\'a}pics et al. (2012) and Kurtz et al. (2015), we examined the extracted frequencies in the form of $f_i = mf_j + nf_k$ to discern potential linear combinations. A peak was identified as a combination frequency if its amplitudes were lower than those of its parent frequencies, and the deviation between the predicted frequency and the observed frequency was less than the frequency resolution of 1.5/$\Delta$T (Loumos $\&$ Deeming 1978). Furthermore, higher-order combinations were considered valid only upon the detection of their lower-order counterparts (P{\'a}pics et al., 2012). Finally, 31 combination frequencies were recognized, leaving 20 confident independent frequencies for further study.

\section{Mode Identifications}
 
According to the theory of stellar oscillations, the rotation will cause frequency splitting, i.e., a single nonradial mode splits into 2$\ell$+1 distinct frequencies. The rotational splitting is generally expressed as 
\begin{equation}
\nu_{\ell,n,m}=\nu_{\ell, n}+m\delta\nu_{\ell,n} = \nu_{\ell,n} + \beta_{\ell,n}\frac{m}{P_{\rm rot}}
\end{equation}
(Saio 1981; Dziembowski \& Goode 1992; Aerts et al. 2010). In the equation, $\ell$, $n$, and $m$ are three indices delineating the oscillation modes, and $\delta\nu_{\ell,n}$ is the splitting frequency. In the case of p modes with high-degree or high-order, $\beta_{\ell,n}$ is approximately equal to 1. The values of rotational splitting for pulsation modes with different $\ell$ are identical. For high-order g modes, $\beta_{\ell,n}$ approximates to $1-\frac{1}{\ell(\ell+1)}$ (Brickhill 1975). The splitting frequency arising from modes with different $\ell$ adheres to a particular proportion (e.g. $\delta\nu_{\ell=1,n} : \delta\nu_{\ell=2,n}$ = 0.6 : 1).

Figure 3 presents results of the Kolmogorov-Smirnov (KS) test (Kawaler 1988) applied to the 20 independent frequencies to identify uniform frequency spacings. The minimal logarithmic value of Q signifies the signicance of a uniform frequency spacing. There are several additional techniques for uncovering regularities in the complex frequency spectra of pulsating stars. Pamos Ortega et al. (2022, 2023) adopted four independent methods to estimate large separation for $\delta$ Scuti stars, including the discrete Fourier transform, the autocorrelation diagram, the frequency difference histogram, and the $\acute{\rm e}$chelle diagram. Su \& Li (2023) employ the inverse variance diagram and the Fourier transform to find the mean period spacing of g-modes for the white dwarf J111215.82 + 111745.0, in addition to the KS test. The results of the different methods are found to be consistent. Breger (2000) demonstrated that HADS stars generally exhibit slow rotation, with $\upsilon\sin$i $\le$ 30 km/s. In Figure 3, a pronounce and significant frequency spacing appears at 2.4 $\mu$Hz, which may correspond to rotational splittings. We examine those frequencies and identify six pairs of frequencies with $\delta\nu$ approximately 2.4 $\mu$Hz, and two pairs of frequencies double the spacing. Table 2 lists details of the possible rotational splittings.

As illustrated in Table 2, the eight pairs of splitting frequencies can be categorized into five groups, including a quintuplet and four sets of doublets. In the table, we can see that there are slight differences for values of rotational splittings in those multiplets. The frequency resolution 1/$\Delta$T is 0.23 $\mu$Hz, which may be the reason for this difference. Moreover, it is worth noting that rotational splittings in different groups are nearly identical, consistent with the characteristic behavior of p modes. For high-amplitude $\delta$ Scuti stars, the dominant peak is commonly inferred to be the radial fundamental mode. Barac et al. (2022) revealed that the dominant peak adheres to the period-luminosity relation. Besides, the period ratio between the dominant peak $f_1$ and $f_3$ is 0.775, in good accordance with the findings of Petersen \& Christensen-Dalsgaard (1996). Consequently, $f_1$ and $f_3$ are proposed to be the radial fundamental mode and the radial first overtone mode, respectively. Given that $f_1$ at 20.4924 c/d, referencing Figure 4 of Chen et al. (2024), TIC 120857354 is suggested to be a young main sequence star, which further confirms the rotational splitting characteristics of p modes.

The average frequency splitting of these five groups is approximately 2.44 $\mu$Hz, which corresponds to a rotation period of 4.75 days. In the table, the frequency discrepancy between $f_7$ and $f_{13}$, and that between $f_{46}$ and $f_{43}$, approximates the averaged value, thereby possessing neighboring azimuthal number, i.e., $\Delta m$ = $\pm$ 1. The frequency difference between $f_4$ and $f_{10}$, as well as between $f_{41}$ and $f_{38}$, approximates twice the averaged value, indicating a separation in azimuthal number by two, i.e., $\Delta m$ = $\pm$ 2. As a result of geometrical cancellation, detections of oscillation modes with higher spherical harmonic degree $\ell$ necessitate substantially more precise photometric data. In this work, we suggest that the quintuplet ($f_{47}$, $f_{20}$, $f_{37}$, $f_{39}$, and $f_{15}$) corresponds to modes of $\ell$ = 2.

From Shibahashi (1979), the asymptotic theory predicts a nearly uniform frequency spacing between high-order p modes with the same $\ell$ and adjacent radial order $n$, i.e.,
\begin{equation}
\Delta \nu = \nu_{\ell, n+1} - \nu_{\ell, n} \approx (2\int_0^R\frac{dr}{c(r)})^{-1}.
\end{equation}
Here, $\Delta\nu$ is the known large frequency separation, and c(r) represents the sound speed. The oscillation modes of $\delta$ Scuti stars are low radial order, they may depart from the asymptotic relation. Murphy et al. (2021) posited that solving the oscillation equations directly may be preferable to gain values of $\Delta\nu$ for $\delta$ Scuti stars. Bedding et al. (2020) demonstrated that frequency spacings of $\ell = 0$ modes decrease as the radial order decreases. The expected large frequency separation should be slightly higher than the frequency difference of 68.733 $\mu$Hz between $f_1$ and $f_3$. In Figure 3, a notable minimum value of $\log Q$ appears at the frequency spacing of $\Delta \nu$ = 37.2 $\mu$Hz, coupled with a localized minimum at twice this value, which may indicate the expected large frequency separation around 74.6 $\mu$Hz. Half the large separations can frequently be found in other approaches of uncovering regularities (Pamos Ortega et al. 2022, 2023). Pamos Ortega et al. (2022, 2023) adopted the most common value derived from various diagnostic techniques as the large separation. In this work, the frequency difference between $f_1$ and $f_3$ can aid us in distinguishing them. The separation is much larger than the value of rotational splitting 2.44 $\mu$Hz. Stellar rotation does not disrupt the frequency spacing pattern, but instead, can aid in mode identification. 

In Group 1, the frequencies ($f_{47}$, $f_{20}$, $f_{37}$, $f_{39}$, and $f_{15}$) are identified as a quintuplet with $\ell$ = 2. Their azimuthal numbers can be uniquely determined as m = (-2, -1, 0, +1, +2). 

For Group 4, the frequency difference between the doublet ($f_{46}$, $f_{43}$) and $f_{37}$ is around 35 $\mu$Hz, which deviates significantly from 74.6 $\mu$Hz. Because $f_{37}$ is already identified as $\ell$ = 2 mode, the doublet ($f_{46}$, $f_{43}$) can only be identified as the modes with $\ell$ =1, while the azimuthal numbers could either be m = (-1, 0) or m = (0, +1).

For Group 5, the frequency difference between the doublet ($f_{41}$, $f_{38}$) and $f_{37}$ is around 72.288 $\mu$Hz, which is close to the expected large frequency separation. Moreover, the frequency differences between the doublets ($f_{41}$, $f_{38}$) and the doublet ($f_{46}$, $f_{43}$) are round 110 $\mu$Hz, significantly greater than expected large frequency separation. Therefore, we propose that frequencies the doublet ($f_{41}$, $f_{38}$) correspond to modes with $\ell$ = 2. There are three possible identifications for their azimuthal numbers: m = (-2, 0), m = (-1, +1), or (0, +2).

Frequency difference between the doublet ($f_7$, $f_{13}$) and the doublet ($f_{46}$, $f_{43}$) is around 74.04 $\mu$Hz. Frequency difference between the doublet ($f_4$, $f_{10}$) and the doublet ($f_{46}$, $f_{43}$) is around 83 $\mu$Hz. The former value is closer to the expected large frequency separation. Thus, we suggest that ($f_7$, $f_{13}$) corresponds to modes with $\ell$= 1, while ($f_4$, $f_{10}$) corresponds to modes with $\ell$ = 2. The azimuthal numbers of ($f_7$, $f_{13}$) could either be m = (-1, 0) or m = (0, +1), while the azimuthal numbers of ($f_4$, $f_{10}$) could be m = (-2, 0), m = (-1, +1), or (0, +2).

Frequencies $f_2$, $f_{21}$, $f_{28}$, $f_{32}$, and $f_{40}$ do not show the rotational splitting effect. The discrepancy in frequency between $f_{28}$ and $f_{32}$ is 71.369 $\mu$Hz, and the frequency difference between $f_{28}$ and $f_{3}$ is 136.87 $\mu$Hz. Therefore, we propose that $f_{28}$ corresponds to the radial third overtone mode, and $f_{32}$ to the radial fourth overtone mode. The difference in frequency between $f_2$ and the doublet ($f_7$, $f_{13}$) is around 75.909 $\mu$Hz, and the frequency difference between $f_{21}$ and the doublet ($f_{46}$, $f_{43}$) is about 74.611 $\mu$Hz. As the mentioned two doublets are modes with l=1, we thus suggest that $f_2$ and $f_{21}$ correspond to two modes with $\ell$ = 1. Nevertheless, their azimuthal numbers remain unidentified. For $f_{40}$, the frequency difference with $f_{37}$ is 64.858 $\mu$Hz, implying $f_{40}$ is a mode with $\ell$ = 2, with its azimuthal number also yet to be identified.

Based on the above analyses, we have identified 4 radial modes, 6 dipole modes, and 10 quadrupole modes. The results are summarized in Table 3 and visually depicted through the frequency $\acute{\rm e}$chelle diagram showcased in Figure 4. Regular frequency sequences corresponding to $\ell$ = 0, $\ell$ = 1, and $\ell$ = 2 modes are distinctly observable in the figure. The sequences for $\ell$ = 0 display a rightward curvature as the radial order decreases, indicating a reduction in $\Delta\nu$ for lower-order modes, whereas the $\ell$ = 1 sequence does not show this characteristic. This feature aligns with the results of Bedding et al. (2020) and Murphy et al. (2022, 2023). In addition, we found that the sequences for $\ell$ = 2 also demonstrate a rightward curvature as the radial order decreases, similar to the $\ell$ = 0 sequence. Moreover, the observation of strictly equidistant frequency intervals in the $\ell =1$ sequence may suggest that $f_2$ and $f_{21}$ has the same azimuthal order with $f_7$ and $f_{46}$. 

\section{Asteroseismic models}
In this work, stellar models were generated with the one-dimensional stellar evolution code known as Modules for Experiments in Stellar Astrophysics (MESA, version 24.08.1, Paxton et al. 2011, 2013, 2015, 2018, 2019, Jermyn et al. 2023). Adiabatic frequencies were computed using the stellar oscillation code GYRE (version 7.2.1, Townsend \& Teitler 2013;Townsend et al. 2018; Goldstein \& Townsend 2020; Sun et al. 2023). A liner grid scan was conducted over 100 - 1000 $\mu$Hz with parameters w\_osc = 10, w\_exp = 5 and w\_ctr =10. The initial stellar abundances followed Asplund et al. (2009), and the corresponding A09 opacity tables were adopted. Default settings were chosen for the equation of state, and the "pp\_cno\_extras\_o18\_ne22.net" nuclear reaction network was employed. The Eddington T-tau relation was applied to model atmospheric boundary conditions and the classical mixing length theory (B$\ddot{\rm o}$hm-Vitense 1958) was engaged to simulate convective. Our models did not include the influences of element diffusion, rotation, and magnetic fields on the structure and evolution of the stars. More details can refer to the inlist in the Appendix. 

Each star in this study evolves from the pre-main-sequence phase to the evolutionary stage where the hydrogen in the core is exhausted ($X_{\rm c}$ = 1$\times$ 10$^{-5}$). The pre-main sequence models were computed with an initial center temperature of 3 $\times$ 10$^5$ K, which signifies the zero-age point for our models. Moreover, we controled the timestep of the star's evolution by setting the parameters time\_delta\_coeff = 0.20 and max\_years\_for\_timestep = 5 $\times$ 10$^5$ years, thereby ensuring that each evolutionary track encompasses approximately 5000 samples. The initial stellar mass spans from 1.4 to 2.2 $M_{\odot}$ with a increment of $\Delta M$  = 0.02 $M_{\odot}$. The initial metallicity $Z$ varies between 0.001 and 0.030 with a increment of $\Delta Z$ = 0.001. The initial helium abundance $Y$ is related to $Z$ through the formula $Y$ = 0.249 + 1.5 $Z$, following Choi et al. (2016).  For the mixing length parameter $\alpha_{\rm MLT}$,  Murphy et al. (2021) found that different values of $\alpha_{\rm MLT}$  have little influence on pre-main-sequence stars, and Joyce \& Tayar (2023) demonstrated that variations of $\alpha_{\rm MLT}$ yield negligible effects for main-sequence stars exceeding a mass of approximately 1.2 $M_{\odot}$. Thus we fixed $\alpha_{\rm MLT}$ to the solar value of 1.90 (Paxton et al. 2013) for all our evolutionary models in this work. The convective core overshoot plays a pivotal role in the structure and evolution of main-sequence stars, since overshooting extends the mixing range of chemical elements and in turn supplies more nuclear fuel to central hydrogen-burning region. However, the extent of convective overshoot mixing remains uncertain. Guo \& Li (2019) analyzed the structure of the convective overshoot within stars of 1.0 - 2.0 $M_{\odot}$, and proposed that the exponential overshooting parameter $f_{\rm ov}$ (Freytag et al. 1996; Herwig (2000) ought to be approximately 0.01 for those stars. In this study, we thus adopt $f_{\rm ov}$ = 0.01 for overshooting at the top of the convective core. The effective temperatures of $\delta$ Scuti stars normally vary between 6000 K to 9000 K. Thus, we calculate adiabatic frequencies of oscillation modes with $\ell$ = 0, 1, and 2 for all evolutionary models that fall inside this range. 

We compare model frequencies with identified modes in Table 3 according to 
\begin{equation}
S_\nu^{2}=\frac{1}{k}\sum(|\nu_{\rm mod,i}-\nu_{\rm obs, i}|^{2}).
\end{equation}
A smaller value of $S_{\nu}^{2}$ signifies a higher likelihood of matching the observations. Here, $k$ = 20, and $\nu_{\rm mod,i}$ and $\nu_{\rm obs,i}$ are a pair of matched model and observed frequencies. When doing model fittings, $m \ne 0$ modes are deduced from $m=0$ modes according to Equation (1), and $\beta_{\ell,n}$ values are determined based on distribution of the radial displacement and the horizontal displacement in the star.

Figure 5 illustrates a color diagram depicting the changes of $S_{\nu}^{2}$ versus the stellar mass $M$ and metallicity $Z$. In the figure, it can be seen that the values of $M$ and $Z$ converge to $M$ = 1.54 $M_{\odot}$ and $Z$ = 0.015. 

Figure 6 displays changes of $S_{\nu}^{2}$ versus the gravitational acceleration $\log g$ and the effective temperature $T_{\rm eff}$. The values of $\log g$ and $T_{\rm eff}$ converge to $\log g$ = 4.27 and $T_{\rm eff}$ = 7441 K, respectively. The TESS input catalog (Stassun et al. 2019) provides $T_{\rm eff}$ and $\log g$ values of 4.42 dex and 8565 K, respectively. While Gaia DR3 (Gaia Collaboration 2022) offers the values of 4.23 for $\log g$ and 7960 K for $T_{\rm eff}$. Our asteroseismic results yield a significantly lower value for $T_{\rm eff}$.

Figure 7 shows the changes of $S_{\nu}^{2}$ versus the stellar age and the relative central hydrogen abundance $X_c/X_0$, where $X_0$ represents the initial hydrogen abundance of the star. In the figure, $X_c/X_0$ = 1 denotes the zero-age main-sequence stage, while $X_c/X_0$ = 0 corresponds to the terminal age of the main-sequence stage. It can be seen that TIC 120857354 is still a young main-sequence star, with $X_c/X_0$ converges to = 0.84, only a small fraction of the initial hydrogen has been burned. The age of the star is estimated to be about 0.53 Gyr.

Table 4 presents fundamental parameters derived by asteroseismology. In the table, the uncertainties are estimated referencing the studies of Steindl et al. (2022) and Su \& Li (2023). We introduced perturbations to the 20 independent frequencies using a uniformly selected random value within -$1/ \Delta T$ and $1/ \Delta T$, where $\Delta T$ signifies the duration of observation. Then we use the theoretical models to match those perturbed frequencies, thereby obtaining parameters that minimize $S_{\nu}^2$. In this work, we repeated this procedure 1000 times, and employed the maximum deviation from the parameters as their corresponding uncertainties. Based on the best-fitting model $M$ = 1.54 $M_{\odot}$ and $Z$ = 0.015, we compare model frequencies with those 20 observed modes of Table 3. The model frequency of $m\ne0$ is deduced from its $m=0$ modes in Table 5 according to Equation (1). The comparison results are depicted in Figure 8 in the form of the frequency $\acute{\rm e}$chelle diagram, and details are listed in Table 6. In the figure, circles denotes modes with $\ell$ = 0, triangles denotes modes with $\ell$ = 1, and squares denotes modes with $\ell$ = 2. Solid symbols correspond to observations, while open symbol are their model counterparts. 

\section{Summary and Discussion}
In this work, we have investigated the pulsational characteristics of the HADS star TIC 120857354 based on the 2-minute cadence TESS photometric data. After removing possible combination frequencies, we obtained 20 confident independent frequencies. Among them, we identify five sets of rotation splittings, including a quintuplet ($f_{47}$, $f_{20}$, $f_{37}$, $f_{39}$, and $f_{15}$) and four pairs of doublets ($f_4$, $f_{10}$), ($f_7$, $f_{13}$), ($f_{46}$, $f_{43}$), and ($f_{41}$, $f_{38}$). The values of splitting frequencies in different groups are approximately identical, reflective of p-mode characteristics. 

The KS test reveals that the anticipated large frequency separation is approximately 74.6 $\mu$Hz. Based on this value and the identified sets of rotational splittings, we finally identified 4 radial modes, 6 dipole modes, and 10 quadrupole modes, corresponding to three sequences in the frequency $\acute{\rm e}$chelle diagram. The frequency separations within the $\ell$ = 0 and 2 sequences are found to decrease towards lower-order modes, whereas the $\ell$ = 1 sequence does not display this trend.

We computed a grid of theoretical models to match the identified frequencies and determined that TIC 120857354 is a young main-sequence star with $M$ = 1.54 $\pm$ 0.04 $M_{\odot}$, $Z$ = 0.015 $\pm$ 0.003, $T_{\rm eff}$ = 7441 $\pm$ 370 K, $\log g$ = 4.27 $\pm$ 0.01, $R$ = 1.52 $\pm$ 0.01 $R_{\odot}$, $L$ = 6.33 $\pm$ 1.53 $L_{\odot}$, age = 0.53 $\pm$ 0.07 Gyr, and $X_c/X_0$ = 0.84 $\pm$ 0.04. In recent years, young $\delta$ Scuti stars have been successful modeled, including both pre-main-sequence stars (Murphy et al. 2021; Steindl et al. 2021a, 2021b, 2022) and main-sequence stars in young clusters, such as the Pleiades (Murphy et al. 2022), the $\alpha$ Per (Pamos Ortega et al. 2022), the Trumpler 10, and Praesepe (Pamos Ortega et al. 2023). Our asteroseisimic results reveal the age of TIC 120857354 is about 0.53 Gyr, while its central hydrogen abundance has only been depleted by about 16\%, signifying that TIC 120857354 is still very young. Given that pressure waves mainly propagate in the star's envelope, it is quite interesting that parameters of TIC 120857354 converge toward a small parameter range. To elucidate this, more comprehensive analysis of the propagation characteristics of oscillation modes within the star is essential.

Figure 9 presents the profiles of Brunt$-$V$\ddot{\rm a}$is$\ddot{\rm a}$l$\ddot{\rm a}$ frequency $N$, the characteristic acoustic frequencies $S_{\ell}$ for $\ell$ = 1 and $\ell$ = 2, as well as the hydrogen abundance inside the best-fitting model, where $S_{\ell}$ = $\sqrt{\ell(\ell+1)}c(r)/r$. Therein, $c(r)$ is the adiabatic sound speed and r is the distance to the center of star. The gray zone in the figure corresponds to the observed frequency range of TIC 120857354, spanning from 237.180 $\mu$Hz to 514.152 $\mu$Hz. Figure 10 displays the profiles of the scaled kinetic energy weight function. The figure reveals that modes with $\ell$ = 0 and $\ell$ = 1 are p-modes, whereas the $\ell = 2$ modes, particularly for low-order modes, are p-dominated mixed modes with pronounced g-mode feature in the propagation zones of g modes. According to theory of stellar oscillations, oscillation modes propagate inside the zones where $\omega$ $>$ $N$, $S_{\ell}$ or $\omega$ $<$ $N$, $S_{\ell}$ and are evanescent elsewhere. Specifically, the regions where $\omega$ $>$ $N$ and $\omega$ $>$ $S_{\ell}$ are the p-mode propagation zones, the regions with $\omega$ $<$ $N$ and $\omega$ $<$ $S_{\ell}$ represents the g-mode propagation zones, and the regions where $S_{\ell}$ $<$ $\omega$ $<$ $N$ or $N$ $<$ $\omega$ $<$  $S_{\ell}$ are are known as the evanescent zones with exponentially decaying behaviour. In Figure 9, it is apparent that the propagation zones of g modes with $\ell$ = 2 are significantly thicker compared to those with $\ell$ = 1 in the observed frequency range, especially for frequencies below 400 $\mu$Hz. Conversely, the evanescent zones between the propagation zones of p modes and g modes for $\ell$ = 2 are markedly thinner compared to those for $\ell$ = 1. The propagation zones of p modes and g modes for $\ell$ = 2 are less distinctly separated compared to those for $\ell$ = 1. A similar feature has been found in the pre-main-sequence star HIP 80088 (Chen \& Li 2018). Following Deheuvels et al. (2012) and Goupil et al. (2013), we use the ratio of mode inertia in the g-mode propagation region $E_g$ over the total mode inertia $E$ to quantify mixed extend of the modes. In Table 5, we noticed that $\ell$ = 2 modes are more prone to exhibiting mixed-mode behavior cpmpared to $\ell$ = 1 modes. We noticed that the frequency sequences of Bedding et al. (2020) mainly contain $\ell$ = 0 and $\ell$ = 1 modes, whereas higher-degree sequences, such as $\ell$ = 2, are often absent. The mixed feature may be one reason that $\ell$ = 2 sequences are often absent. P modes mainly propagate in the envelope of the star, while the mixed modes enable us probe deeper into interiors of the star. Following Chen et al. (2024), we use the stellar mean density $\bar{\rho}/\bar{\rho}_{\odot}$ to characterize features of the stellar envelope and use the relative radius of the convective core $R_c$/$R$ to characterize features of the deep interior of the star. Figure 11 illustrates changes of $S_\nu^2$ as a function of $R_c$/$R$ and $\bar{\rho}/\bar{\rho}_{\odot}$. In the figure, $\bar{\rho}/\bar{\rho}_{\odot}$ converges toward 0.444 and $R_c$/$R$ converges toward 0.092.

The average frequency splitting is estimated to be 2.44 $\mu$Hz, corresponding to a rotation period of 4.75 days. Ouazzani et al. (2010) examined the rotational splittings of $\beta$ Cephei stars and demonstrated that the validity threshold of perturbative methods extends to 10\% of the break-up velocity. The break-up velocity of TIC 120857354 is estimated to be 441 km s$^{-1}$. The rotation velocity is inferred to be 16 km s$^{-1}$ through $\upsilon_{\rm e}$=$2\pi R/P_{\rm rot}$, which approximates 3.6\% of the break-up velocity. As a result, the linear perturbation method remains valid. Moreover, considering its low rotational velocity, the influences of rotation on the stellar structure and evolution are not included in our calculations. According to the studies of Saio (1981), Dziembowski \& Goode (1992), and Aerts et al. (2010), the first-order rotational effect on pulsation is proportional to $1/P_{\rm rot}$, and the second-order effect is proportional to $1/(P_{\rm rot}^2\nu_{\ell, n})$. The ratio of the second-order to the first-order rotational effect can be estimated to be on the order of $1/(P_{\rm rot}\nu_{\ell, n})$, wherein $P_{\rm rot}$ is 4.75 days and $\nu_{\ell, n}$ ranges from 306.561 $\mu$Hz to 505.863 $\mu$Hz. The second-order effect of rotation is significantly less than the first-order effect. Therefore, the second-order effect of rotation on pulsation is not considered in this study.

Finally, the asteroseismic determination of $T_{\rm eff}$ is significantly lower than the values of the TESS input catalog and Gaia DR3. Future spectroscopic investigations of TIC 120857354 may be warranted. Moreover, for young $\delta$ Scuti stars, modes with $\ell$ = 2, which are more prone to display mixed-mode behavior, might play a more crucial role than previously thought in uncovering the star's evolutionary history and internal structure. More such objects merit further investigation in the future.

\section*{Acknowledgements}
We are sincerely grateful to the anonymous referee for instructive advice and productive suggestions. The authors acknowledge the supports from the National Natural Science Foundation of China (Grant No. 12288102), the National Key R\&D Program of China (Grant No 2021YFA1600400/2021YFA1600402) and the B-type Strategic Priority Program No.XDB41000000 funded by the Chinese Academy of Sciences. The authors also acknowledge the support of the National Natural Science Foundation of China (12173080, 12133011, 12433009, and 12373037) and the science research grants from the China Manned Space Project. X-H.C. also sincerely  appreciates the supports of the Yunnan Revitalization Talent Support Program Young Talent Project, and the Youth Innovation Promotion Association of the Chinese Academy of Sciences,  as well as the International Centre of Supernovae Yunnan Key Laboratory (No. 202302AN360001). This paper includes data collected with the TESS mission, obtained from the MAST data archive at the Space Telescope Science Institute (STScI). Funding for the TESS mission is provided by the NASA Explorer Program. STScI is operated by the Association of Universities for Research in Astronomy, Inc., under NASA contract NAS 5-26555. The authors sincerely acknowledge them for providing such excellent data.

\section*{Data  availability}

The data that support the findings of this study are available from the corresponding author upon reasonable request.

\appendix 
\section{Inlist files used in this work (Version 24.08.1)}
\begin{verbatim}
! inlist_gyre24.08.1
&star_job
 astero_just_call_my_extras_check_model = .true.
 show_log_description_at_start = .false.
 create_pre_main_sequence_model = .true.
 change_initial_net = .true.
 new_net_name = 'pp_cno_extras_o18_ne22.net'
 initial_zfracs = 6
 set_initial_age = .true.
 initial_age = 0 ! in years
 set_initial_model_number = .true.
 initial_model_number = 0
 pre_ms_relax_num_steps = 100
 pre_ms_relax_to_start_radiative_core = .false.
/ ! end of star_job namelist
&eos
/ ! end of eos namelist
&kap
 use_Zbase_for_Type1 = .false.
 use_Type2_opacities = .false.
 kap_file_prefix = 'a09'
 kap_lowT_prefix = 'lowT_fa05_a09p'
/ ! end of kap namelist
&controls
 energy_eqn_option = 'dedt'
 use_gold2_tolerances = .true.
 max_model_number = -1
 initial_mass = 1.54
 initial_z  =  0.015
 initial_y  =  0.2715
 MLT_option = 'ML1'  
 mixing_length_alpha = 1.90
 min_overshoot_q = 1d-3
 overshoot_D_min = 1d-2
 overshoot_scheme(1) = 'exponential'
 overshoot_zone_type(1) = 'burn_H'
 overshoot_zone_loc(1) = 'core'
 overshoot_bdy_loc(1) = 'top'
 overshoot_f0(1) = 0.005
 overshoot_f(1) =  0.010
 atm_option = 'T_tau'
 atm_T_tau_relation = 'Eddington'
 atm_T_tau_opacity = 'varying'
 atm_build_tau_outer = 1d-5
 atm_build_dlogtau = 0.01
 atm_build_errtol = 1d-8
 calculate_Brunt_N2 =.true.
 time_delta_coeff = 0.20
 max_years_for_timestep= 5d5
 max_allowed_nz=-1
 use_other_mesh_functions= .true.
 mesh_delta_coeff = 0.50
 max_center_cell_dq = 1d-10
 xa_central_lower_limit_species(1) = 'h1'
 xa_central_lower_limit(1) = 1d-5
/ ! end of controls namelist
\end{verbatim}


\bibliographystyle{mnras}

\begin{thebibliography}{}


\bibitem[Aerts et al.(2010)]{2010aste.book.....A} Aerts, C., Christensen-Dalsgaard, J., \& Kurtz, D.~W.\ 2010, Asteroseismology, Astronomy and Astrophysics Library (Berlin: Springer)

\bibitem[Aerts(2021)]{2021RvMP...93a5001A} Aerts, C.\ 2021, Reviews of Modern Physics, 93, 015001

\bibitem[Antoci et al.(2014)]{2014ApJ...796..118A} Antoci, V., Cunha, M., Houdek, G., et al.\ 2014, \apj, 796, 118

\bibitem[Antoci et al.(2019)]{2019MNRAS.490.4040A} Antoci, V., Cunha, M.~S., Bowman, D.~M., et al.\ 2019, \mnras, 490, 4040

\bibitem[Asplund et al.(2009)]{2009ARA&A..47..481A} Asplund, M., Grevesse, N., Sauval, A.~J., \& Scott, P.\ 2009, ARA\&A, 47, 481

\bibitem[Baglin et al.(2006)]{2006ESASP1306...33B} Baglin, A., Auvergne, M., Barge, P., et al. 2006, in ESA SP 1306, The CoRoT Mission Pre-Launch Status?Stellar Seismology and Planet Finding, ed. M. Fridlund et al. (Paris: ESA), 33

\bibitem [Baker(1962)] {Ba62} Baker, N., $\&$ Kippenhahn, R. 1962, Z. f. Astrophys., 54, 114

\bibitem[Baker(1965)]{Ba65} Baker, N., $\&$ Kippenhahn, R. 1965, APJ, 142, 868

\bibitem[Balona et al.(2012)]{2012MNRAS.419.3028B} Balona, L.~A., Lenz, P., Antoci, V., et al.\ 2012, \mnras, 419, 3028

\bibitem[Barac et al.(2022)]{2022MNRAS.516.2080B} Barac, N., Bedding, T.~R., Murphy, S.~J., et al.\ 2022, \mnras, 516, 2080

\bibitem[Barcel{\'o} Forteza et al.(2015)]{2015A&A...579A.133B} Barcel{\'o} Forteza, S., Michel, E., Roca Cort{\'e}s, T., \& Garc{\'{\i}}a, R.~A.\ 2015, \aap, 579, A133 

\bibitem[Barcel{\'o} Forteza et al.(2018)]{2018A&A...614A..46B} Barcel{\'o} Forteza, S., Roca Cort{\'e}s, T., \& Garc{\'{\i}}a, R.~A.\ 2018, \aap, 614, A46 

\bibitem[Barcel{\'o} Forteza et al.(2020)]{2020A&A...638A..59B} Barcel{\'o} Forteza, S., Moya, A., Barrado, D., et al.\ 2020, \aap, 638, A59


\bibitem[Bedding et al.(2020)]{2020Natur.581..147B} Bedding, T.~R., Murphy, S.~J., Hey, D.~R., et al.\ 2020, Nature, 581, 147

\bibitem[Bedding et al.(2023)]{2023ApJ...946L..10B} Bedding, T.~R., Murphy, S.~J., Crawford, C., et al.\ 2023, \apjl, 946, L10

\bibitem[B{\"o}hm-Vitense(1958)]{1958ZA.....46..108B} B{\"o}hm-Vitense, E.\ 1958, \zap, 46,
 108
 
\bibitem[Borucki et al.(2010)]{2010Sci...327..977B} Borucki, W.~J., Koch, D., Basri, G., et al.\ 2010, Science, 327, 977

\bibitem[Bowman \& Kurtz(2014)]{2014MNRAS.444.1909B} Bowman, D.~M., \& Kurtz, D.~W.\ 2014, \mnras, 444, 1909 

\bibitem[Bowman et al.(2016)]{2016MNRAS.460.1970B} Bowman, D.~M., Kurtz, D.~W., Breger, M., Murphy, S.~J., \& Holdsworth, D.~L.\ 2016, \mnras, 460, 1970 


\bibitem[Breger et al.(1993)]{1993A&A...271..482B} Breger, M., Stich, J., Garrido, R., et al.\ 1993, \aap, 271, 482

\bibitem[Breger(2000)]{2000ASPC..210....3B} Breger, M.\ 2000, Delta Scuti and Related Stars, 210, 3 

\bibitem[Breger et al.(2011)]{2011MNRAS.414.1721B} Breger, M., Balona, L., Lenz, P., et al.\ 2011, \mnras, 414, 1721

\bibitem[Breger et al.(2017)]{2017A&A...599A.116B} Breger, M., Montgomery, M.~H., Lenz, P., et al.\ 2017, \aap, 599, A116

\bibitem[Brickhill(1975)]{1975MNRAS.170..405B} Brickhill, A.~J.\ 1975, \mnras, 170, 405

\bibitem [Castanheira(2008)] {Ca08} Castanheira, B. G., et al. 2008, CoAst, 157, 124

\bibitem[Chen et al.(2016)]{2016A&A...593A..69C} Chen, X.~H., Li, Y., Lai, X.~J., et al.\ 2016, \aap, 593, A69

\bibitem[Chen et al.(2017)]{2017ApJ...834..146C} Chen, X.~H, Li, Y., Lin, G.~F, et al.\ 2017, \apj, 834, 146

\bibitem[Chen \& Li(2017)]{2017ApJ...838...31C} Chen, X.~H \& Li, Y.\ 2017, \apj, 838, 31

\bibitem[Chen \& Li(2018)]{2018ApJ...866..147C} Chen, X.~H \& Li, Y.\ 2018, \apj, 866, 147

\bibitem[Chen et al.(2024)]{2024ApJ...963..155C} Chen, X.~H, Zhang, X.~B, Li, Y., et al.\ 2024, \apj, 963, 155

\bibitem[Chevalier(1971)]{1971A&A....14...24C} Chevalier, C.\ 1971, \aap, 14, 24 

\bibitem[Choi et al.(2016)]{2016ApJ...823..102C} Choi, J., Dotter, A., Conroy, C., et al.\ 2016, \apj, 823, 102

\bibitem[Chang et al.(2013)]{2013AJ....145..132C} Chang, S.-W., Protopapas, P., Kim, D.-W., et al.\ 2013, \aj, 145, 132

\bibitem [Daszynska(2005)] {Da05} Daszy\'{n}ska-Daszkiewicz, J., Dziembowski, W. A., et al. 2005, A$\&$A, 438, 653

\bibitem[Deheuvels et al.(2012)]{2012ApJ...756...19D} Deheuvels, S., Garc{\'\i}a, R.~A., Chaplin, W.~J., et al.\ 2012, \apj, 756, 19

\bibitem[Dupret et al.(2004)]{2004A&A...414L..17D} Dupret, M.-A., Grigahc{\`e}ne, A., Garrido, R., Gabriel, M., \& Scuflaire, R.\ 2004, \aap, 414, L17 

\bibitem[Dziembowski \& Goode(1992)]{1992ApJ...394..670D} Dziembowski, W.~A., \& Goode, P.~R.\ 1992, \apj, 394, 670

\bibitem[Freytag et al.(1996)]{1996A&A...313..497F} Freytag, B., Ludwig, H.-G., \& Steffen, M.\ 1996, \aap, 313, 497

\bibitem[Fu et al.(2013)]{2013MNRAS.429.1585F} Fu, J.~N., Dolez, N., Vauclair, G., et al.\ 2013, \mnras, 429, 1585

\bibitem[Gaia Collaboration(2022)]{2022yCat.1355....0G} Gaia Collaboration\ 2022, VizieR Online Data Catalog, 1355

\bibitem[Garc{\'{\i}}a Hern{\'a}ndez et al.(2015)]{2015ApJ...811L..29G} Garc{\'{\i}}a Hern{\'a}ndez, A., Mart{\'{\i}}n-Ruiz, S., Monteiro, M.~J.~P.~F.~G., et al.\ 2015, ApJL, 811, L29 

\bibitem[Garc{\'{\i}}a Hern{\'a}ndez et al.(2017)]{2017MNRAS.471L.140G} Garc{\'{\i}}a Hern{\'a}ndez, A., Su{\'a}rez, J.~C., Moya, A., et al.\ 2017, \mnras, 471, L140 

\bibitem[Goupil et al.(2013)]{2013A&A...549A..75G} Goupil, M.~J., Mosser, B., Marques, J.~P., et al.\ 2013, \aap, 549, A75

\bibitem[Grigahc{\`e}ne et al.(2005)]{2005A&A...434.1055G} Grigahc{\`e}ne, A., Dupret, M.-A., Gabriel, M., Garrido, R., \& Scuflaire, R.\ 2005, \aap, 434, 1055 

\bibitem[Goldstein \& Townsend(2020)]{2020ApJ...899..116G} Goldstein, J. \& Townsend, R.~H.~D.\ 2020, \apj, 899, 116

\bibitem[Guo \& Li(2019)]{2019ApJ...879...86G} Guo, F. \& Li, Y.\ 2019, \apj, 879, 86

\bibitem[Herwig(2000)]{2000A&A...360..952H} Herwig, F.\ 2000, \aap, 360, 952

\bibitem[Houdek(2000)]{2000ASPC..210..454H} Houdek, G.\ 2000, Delta Scuti and Related Stars, 210, 454

\bibitem[Houdek(2008)]{2008CoAst.157..137H} Houdek, G.\ 2008, Communications in Asteroseismology, 157, 137. doi:10.48550/arXiv.0810.5228

\bibitem[Jenkins et al.(2016)]{2016SPIE.9913E..3EJ} Jenkins, J.~M., Twicken, J.~D., McCauliff, S., et al.\ 2016, Society of Photo-Optical Instrumentation Engineers(SPIE) Conference Series, Vol. 9913, The TESS scienceprocessing operations center, 99133E

\bibitem[Jermyn et al.(2023)]{2023ApJS..265...15J} Jermyn, A.~S., Bauer, E.~B., Schwab, J., et al.\ 2023, \apjs, 265, 15

\bibitem[Joyce \& Tayar(2023)]{2023Galax..11...75J} Joyce, M. \& Tayar, J.\ 2023, Galaxies, 11, 75

\bibitem[Kurtz et al.(2014)]{2014MNRAS.444..102K} Kurtz, D.~W., Saio, H., Takata, M., et al.\ 2014, \mnras, 444, 102

\bibitem[Kurtz et al.(2015)]{2015MNRAS.450.3015K} Kurtz, D.~W., Shibahashi, H., Murphy, S.~J., et al.\ 2015, \mnras, 450, 3015

\bibitem[Kawaler(1988)]{1988IAUS..123..329K} Kawaler, S.~D.\ 1988, Advances in Helio- and Asteroseismology, 123, 329

\bibitem[Lenz \& Breger(2005)]{2005CoAst.146...53L} Lenz, P. \& Breger, M.\ 2005, Communications in Asteroseismology, 146, 53

\bibitem[Li et al.(2024)]{2024A&A...686A.142L} Li, G., Aerts, C., Bedding, T.~R., et al.\ 2024, \aap, 686, A142

\bibitem [Li(1994)] {Li94} Li, Y., $\&$ Stix, M. 1994, A$\&$A, 286, 815

\bibitem[Loumos \& Deeming(1978)]{1978Ap&SS..56..285L} Loumos, G.~L., \& Deeming, T.~J.\ 1978, \apss, 56, 285

\bibitem[Mantegazza(2000)]{2000ASPC..210..138M} Mantegazza, L.\ 2000, Delta Scuti and Related Stars, 210, 138

\bibitem[McNamara(2000)]{2000PASP..112.1096M} McNamara, D.~H.\ 2000, PASP, 112, 1096

\bibitem[Murphy et al.(2020)]{2020MNRAS.498.4272M} Murphy, S.~J., Saio, H., Takada-Hidai, M., et al.\ 2020, \mnras, 498, 4272

\bibitem[Murphy et al.(2021)]{2021MNRAS.502.1633M} Murphy, S.~J., Joyce, M., Bedding, T.~R., et al.\ 2021, \mnras, 502, 1633

\bibitem[Murphy et al.(2022)]{2022MNRAS.511.5718M} Murphy, S.~J., Bedding, T.~R., White, T.~R., et al.\ 2022, \mnras, 511, 5718

\bibitem[Murphy et al.(2023)]{2023MNRAS.526.3779M} Murphy, S.~J., Bedding, T.~R., Gautam, A., et al.\ 2023, \mnras, 526, 3779

\bibitem[Murphy et al.(2024)]{2024MNRAS.534.3022M} Murphy, S.~J., Bedding, T.~R., Gautam, A., et al.\ 2024, \mnras, 534, 3022

\bibitem[Ouazzani et al.(2010)]{2010Ap&SS.328..285O} Ouazzani, R.~M., Goupil, M.~J., Dupret, M.~A., et al.\ 2010, \apss, 328, 285

\bibitem[Pamos Ortega et al.(2022)]{2022MNRAS.513..374P} Pamos Ortega, D., Garc{\'\i}a Hern{\'a}ndez, A., Su{\'a}rez, J.~C., et al.\ 2022, \mnras, 513, 374

\bibitem[Pamos Ortega et al.(2023)]{2023A&A...675A.167P} Pamos Ortega, D., Mirouh, G.~M., Garc{\'\i}a Hern{\'a}ndez, A., et al.\ 2023, \aap, 675, A167

\bibitem[Papar{\'o} et al.(2016a)]{2016ApJ...822..100P} Papar{\'o}, M., Benk{\H{o}}, J.~M., Hareter, M., et al.\ 2016a, \apj, 822, 100

\bibitem[Papar{\'o} et al.(2016b)]{2016ApJS..224...41P} Papar{\'o}, M., Benk{\H{o}}, J.~M., Hareter, M., et al.\ 2016b, \apjs, 224, 41

\bibitem[P{\'a}pics(2012)]{2012AN....333.1053P} P{\'a}pics, P.~I.\ 2012, Astronomische Nachrichten, 333, 1053

\bibitem[Paxton et al.(2011)]{2011ApJS..192....3P} Paxton, B., Bildsten, L., Dotter, A., et al.\ 2011, \apjs, 192, 3 

\bibitem[Paxton et al.(2013)]{2013ApJS..208....4P} Paxton, B., Cantiello, M., Arras, P., et al.\ 2013, \apjs, 208, 4 

\bibitem[Paxton et al.(2015)]{2015ApJS..220...15P} Paxton, B., Marchant, P., Schwab, J., et al.\ 2015, \apjs, 220, 15 

\bibitem[Paxton et al.(2018)]{2018ApJS..234...34P} Paxton, B., Schwab, J., Bauer, E.~B., et al.\ 2018, \apjs, 234, 34 

\bibitem[Paxton et al.(2019)]{2019ApJS..243...10P} Paxton, B., Smolec, R., Schwab, J., et al.\ 2019, \apjs, 243, 10

\bibitem[Petersen(1989)]{1989DSSN....1....6P} Petersen, J.~O.\ 1989, Delta Scuti Star Newsletter, vol. 1, p.6, 1

\bibitem[Petersen \& Christensen-Dalsgaard(1996)]{1996A&A...312..463P} Petersen, J.~O. \& Christensen-Dalsgaard, J.\ 1996, \aap, 312, 463

\bibitem[Poretti(2003)]{2003A&A...409.1031P} Poretti, E.\ 2003, \aap, 409, 1031

\bibitem[Poretti et al.(2005)]{2005A&A...440.1097P} Poretti, E., Su{\'a}rez, J.~C., Niarchos, P.~G., et al.\ 2005, \aap, 440, 1097

\bibitem[Poretti et al.(2011)]{2011A&A...528A.147P} Poretti, E., Rainer, M., Weiss, W.~W., et al.\ 2011, \aap, 528, A147

\bibitem[Ricker et al.(2015)]{2015JATIS...1a4003R} Ricker, G.~R., Winn, J.~N., Vanderspek, R., et al.\ 2015, Journal of Astronomical Telescopes, Instruments, and Systems, 1, 014003

\bibitem[Saio(1981)]{1981ApJ...244..299S} Saio, H.\ 1981, \apj, 244, 299 

\bibitem [Schmid(2014)] {Sc14} Schmid, V. S., et al. 2014, A$\&$A, 570, A33

\bibitem[Slawson et al.(2011)]{2011AJ....142..160S} Slawson, R.~W., Pr{\v{s}}a, A., Welsh, W.~F., et al.\ 2011, \aj, 142, 160

\bibitem[Stassun et al.(2019)]{2019AJ....158..138S} Stassun, K.~G., Oelkers, R.~J., Paegert, M., et al.\ 2019, \aj, 158, 138

\bibitem[Steindl et al.(2021a)]{2021A&A...645A.119S} Steindl, T., Zwintz, K., \& Bowman, D.~M.\ 2021a, \aap, 645, A119

\bibitem[Steindl et al.(2021b)]{2021A&A...654A..36S} Steindl, T., Zwintz, K., Barnes, T.~G., et al.\ 2021b, \aap, 654, A36

\bibitem[Steindl et al.(2022)]{2022A&A...664A..32S} Steindl, T., Zwintz, K., \& M{\"u}llner, M.\ 2022, \aap, 664, A32

\bibitem[Stellingwerf(1979)]{1979ApJ...227..935S} Stellingwerf, R.~F.\ 1979, \apj, 227, 935

\bibitem[Su \& Li(2023)]{2023ApJ...943..113S} Su, J. \& Li, Y.\ 2023, \apj, 943, 113

\bibitem[Sun et al.(2023)]{2023ApJ...945...43S} Sun, M., Townsend, R.~H.~D., \& Guo, Z.\ 2023, \apj, 945, 43

\bibitem[Sun et al.(2021)]{2021ApJ...922..199S} Sun, X.-Y., Zuo, Z.-Y., Yang, T.-Z., et al.\ 2021, \apj, 922, 199

\bibitem[Su{\'a}rez et al.(2014)]{2014A&A...563A...7S} Su{\'a}rez, J.~C., Garc{\'{\i}}a Hern{\'a}ndez, A., Moya, A., et al.\ 2014, \aap, 563, A7 

\bibitem[Townsend \& Teitler(2013)]{2013MNRAS.435.3406T} Townsend, R.~H.~D. \& Teitler, S.~A.\ 2013, \mnras, 435, 3406

\bibitem[\protect\citeauthoryear{Townsend, Goldstein, \& Zweibel}{2018}]{2018MNRAS.475..879T} Townsend R.~H.~D., Goldstein J., Zweibel E.~G., 2018, MNRAS, 475, 879

\bibitem[Walraven et al.(1992)]{1992MNRAS.254...59W} Walraven, T., Walraven, J., \& Balona, L.~A.\ 1992, \mnras, 254, 59

\bibitem[Watson(1988)]{1988Ap&SS.140..255W} Watson, R.~D.\ 1988, \apss, 140, 255

\bibitem[Xiong et al.(2016)]{2016MNRAS.457.3163X} Xiong, D.~R., Deng, L., Zhang, C., et al.\ 2016, \mnras, 457, 3163

\bibitem[Xue et al.(2018)]{2018ApJ...861...96X} Xue, H.-F., Fu, J.-N., Fox-Machado, L., et al.\ 2018, \apj, 861, 96

\bibitem[Yang et al.(2021)]{2021A&A...655A..63Y} Yang, T.-Z., Zuo, Z.-Y., Li, G., et al.\ 2021, \aap, 655, A63

\bibitem[Zhevakin (1963)]{Zh63} Zhevakin, S. A. 1963, Ann. Rev. Astr. Astrophys., 1, 367

\bibitem [Zima(2006)] {Zi06} Zima, W., Wright, D., Bentley, J., et al. 2006, A$\&$A, 455, 235

\end{thebibliography}

\begin{figure*}
\centering
\includegraphics[width=\textwidth, angle = 0]{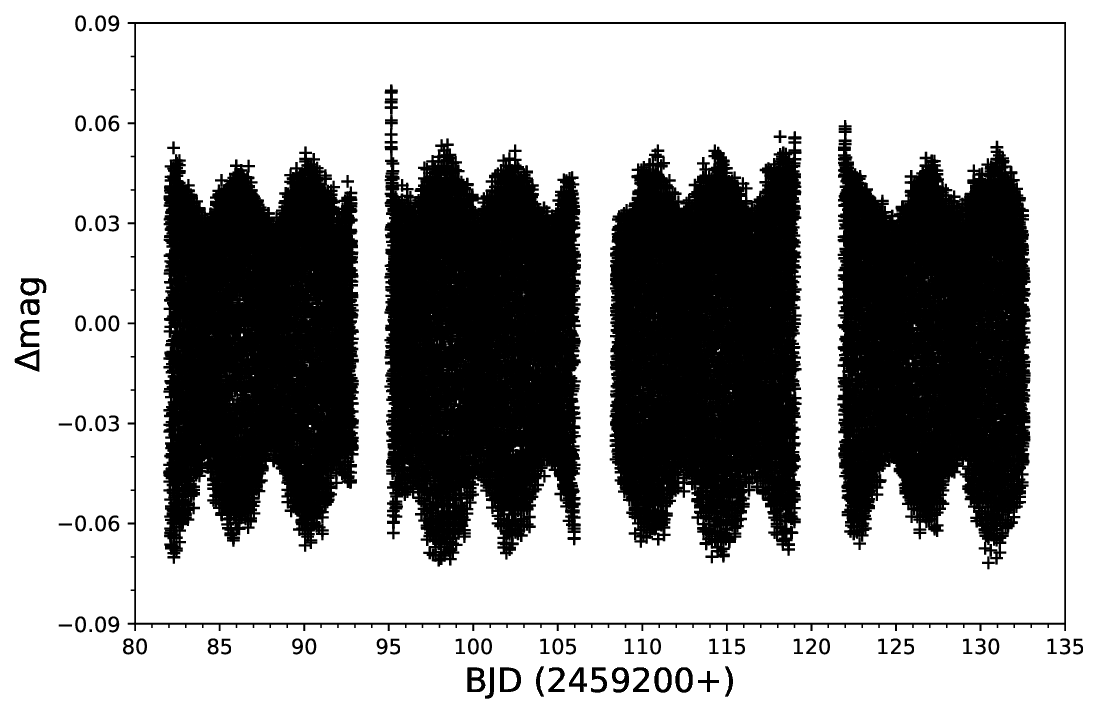}
\caption{\label{Figure 1} Light curve of TIC 120857354 observed by TESS in the 2-minute cadence during Sectors 36 and 37. The amplitude of the light curve is about 0.12 mag. }
\end{figure*}
\begin{figure*}
\centering
\includegraphics[width=\textwidth, angle = 0]{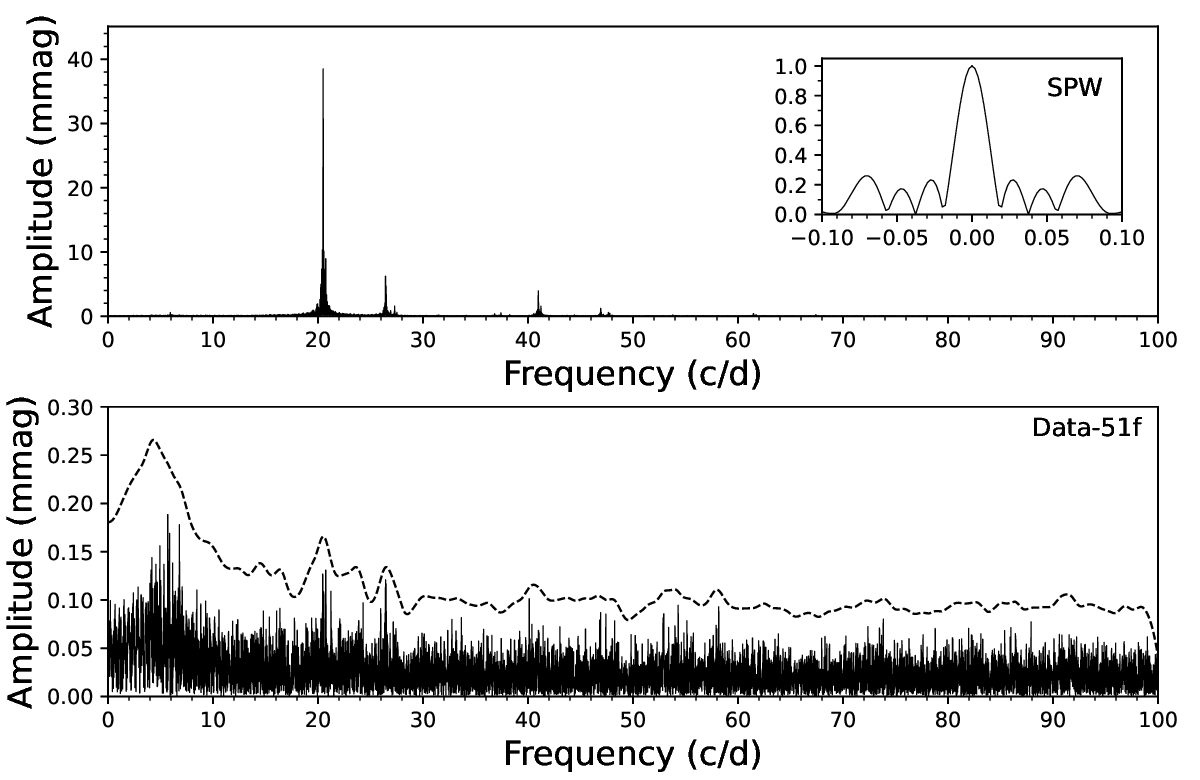}
\caption{\label{Figure 2} Fourier amplitude spectrum. The upper panel presents the original spectrum, in which the inset panel is the window spectrum. The lower panel is the residual spectrum after 51 frequencies with S/N $>$ 4 were subtracted.The dashed line marks the level of S/N = 4.}
\end{figure*}

\begin{figure*}
\centering
\includegraphics[width=\textwidth, angle = 0]{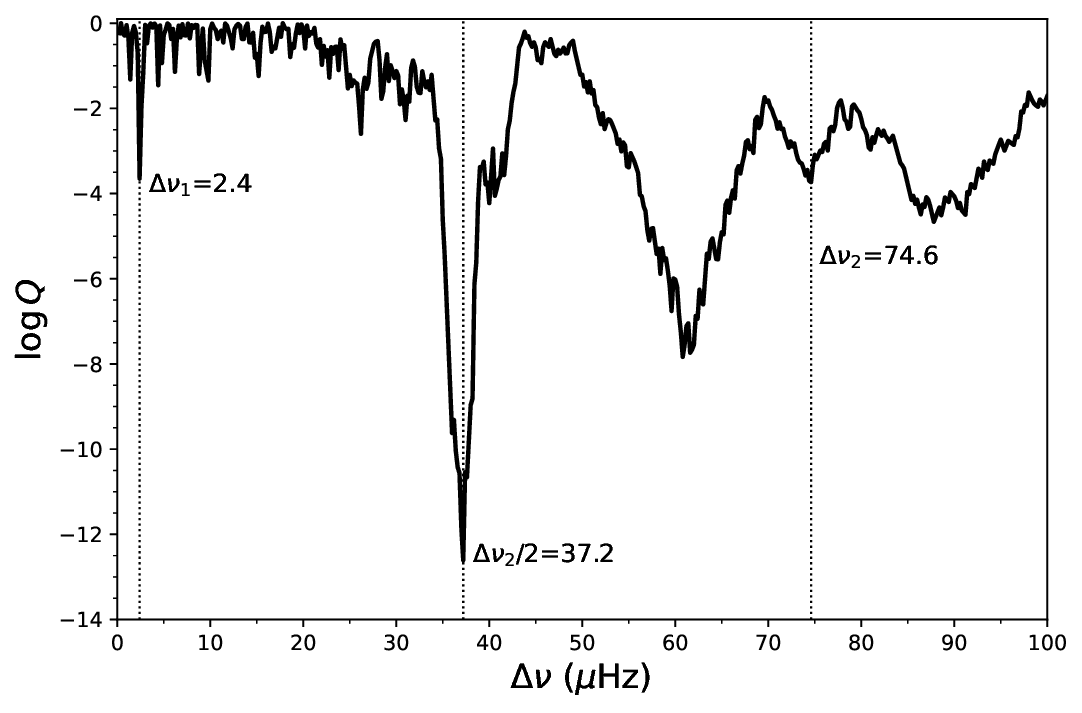}
  \caption{\label{Figure 3} Results of KS test applied to the 20 independent frequencies.}
\end{figure*}

\begin{figure*}
\centering
\includegraphics[width=\textwidth, angle = 0]{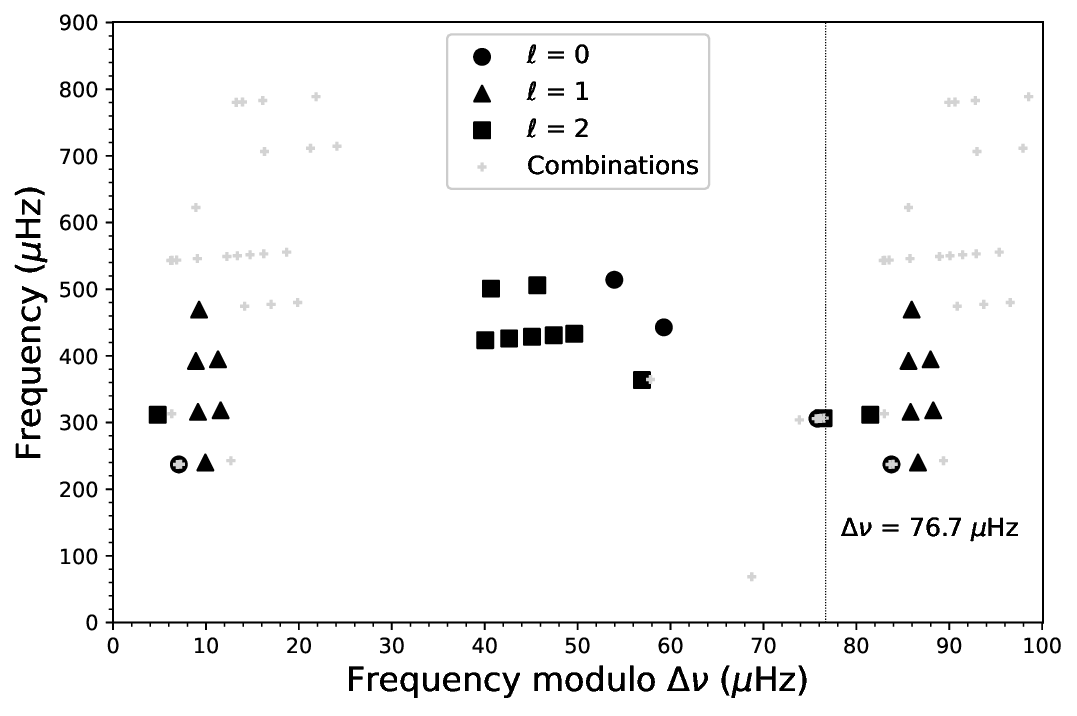}
  \caption{\label{Figure 4} Frequency $\acute{\rm e}$chelle diagram for TIC 120857354, with large frequency separation $\Delta \nu$ = 76.7 $\mu$Hz. Solid circles correspond to the four $\ell$ = 0 in Table 3, solid triangles correspond to six $\ell$ = 1 modes in Table 3, solid squares correspond to ten $\ell$ = 2 modes in table 3, and grey crosses correspond to 31 combinations or unresolved frequencies in Table 1. The $\acute{\rm e}$chelle diagram is repeated to the right of the dashed line to enhance visibility. }
\end{figure*}

\begin{figure*}
\centering
\includegraphics[width=\textwidth, angle = 0]{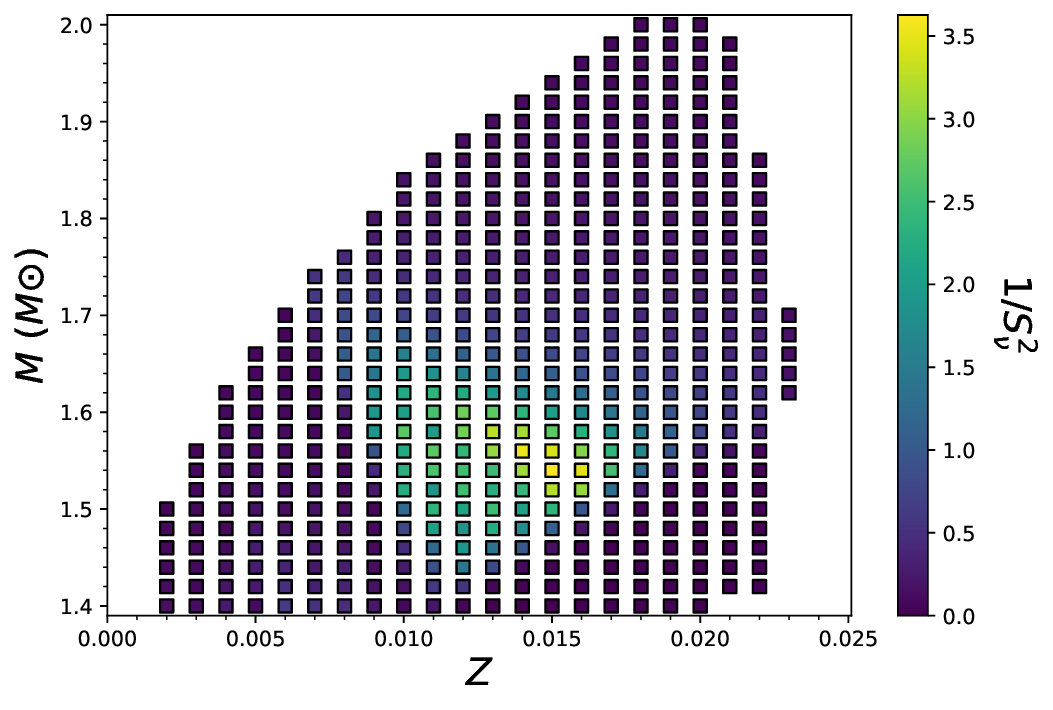}
  \caption{\label{Figure 5}  Color diagram of the fitting results 1/S$_{\nu}^2$ versus stellar mass $M$ and metaillicity $Z$. Each square in the figure represents one
minimum value of $S_{\nu}^2$ along one evolutionary track. }
\end{figure*}

\begin{figure*}
\centering
\includegraphics[width=\textwidth, angle = 0]{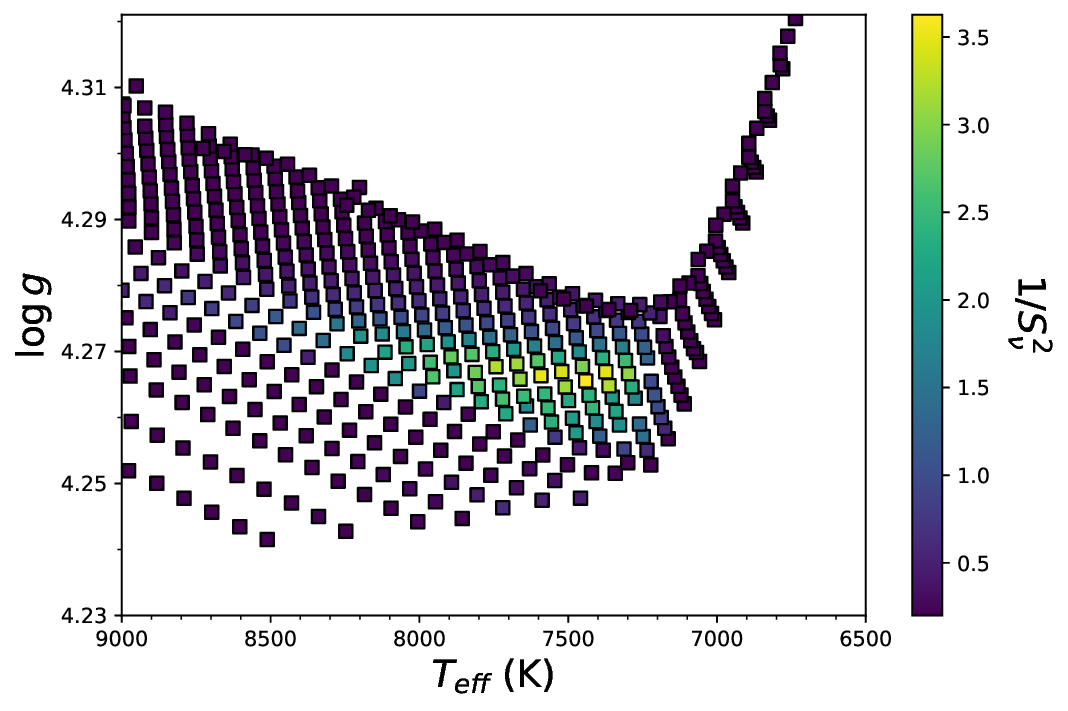}
  \caption{\label{Figure 6} Color diagram of the fitting results 1/S$_{\nu}^2$ versus the gravitational acceleration $\log g$ and the effective temperature $T_{\rm eff}$.}
\end{figure*}

\begin{figure*}
\centering
\includegraphics[width=\textwidth, angle = 0]{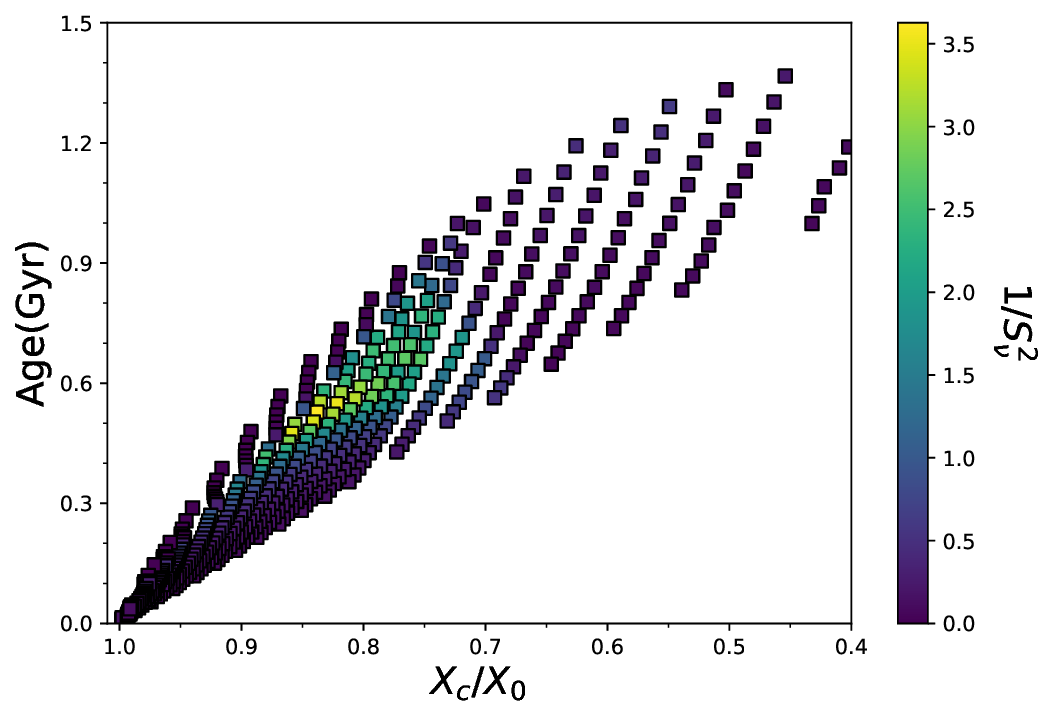}
  \caption{\label{Figure 7}  Color diagram of the fitting results 1/S$_{\nu}^2$ versus stellar age and the normalized central hydrogen abundance $X_c/X_0$. }
\end{figure*}

\begin{figure*}
\centering
\includegraphics[width=\textwidth, angle = 0]{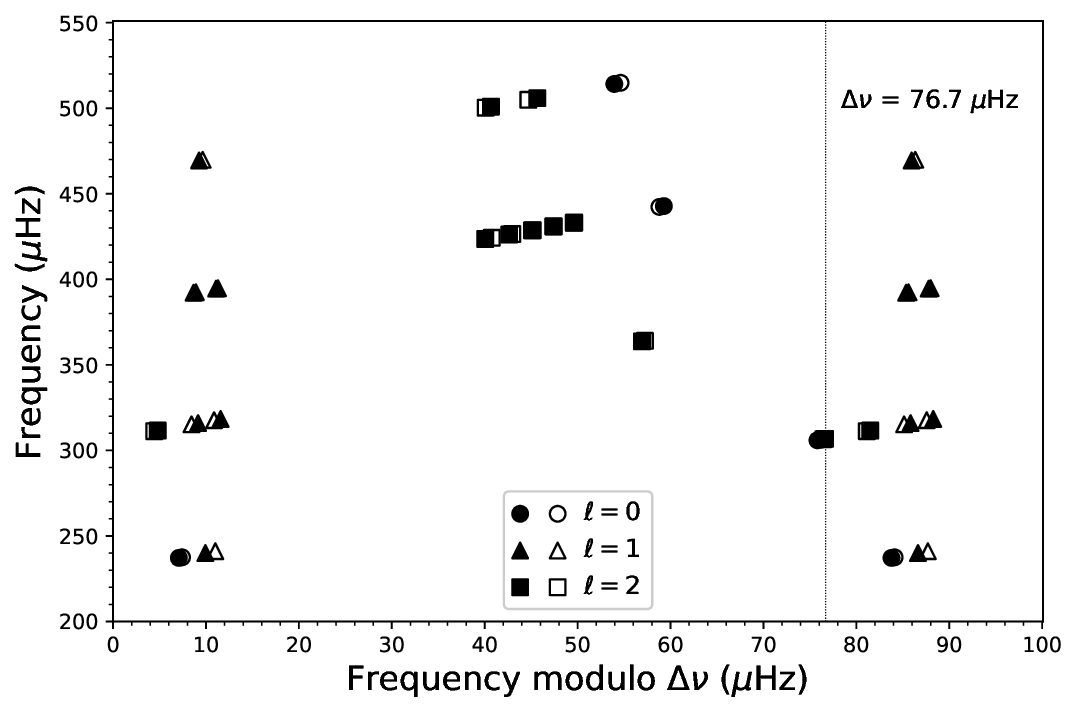}
  \caption{\label{Figure 8} Comparison results between observations and the best-fitting model on the frequency $\acute{\rm e}$chelle diagram. Solid symbols correspond to observations, while open symbol are their model counterparts. }
\end{figure*}

\begin{figure*}
\centering
\includegraphics[width=\textwidth, angle = 0]{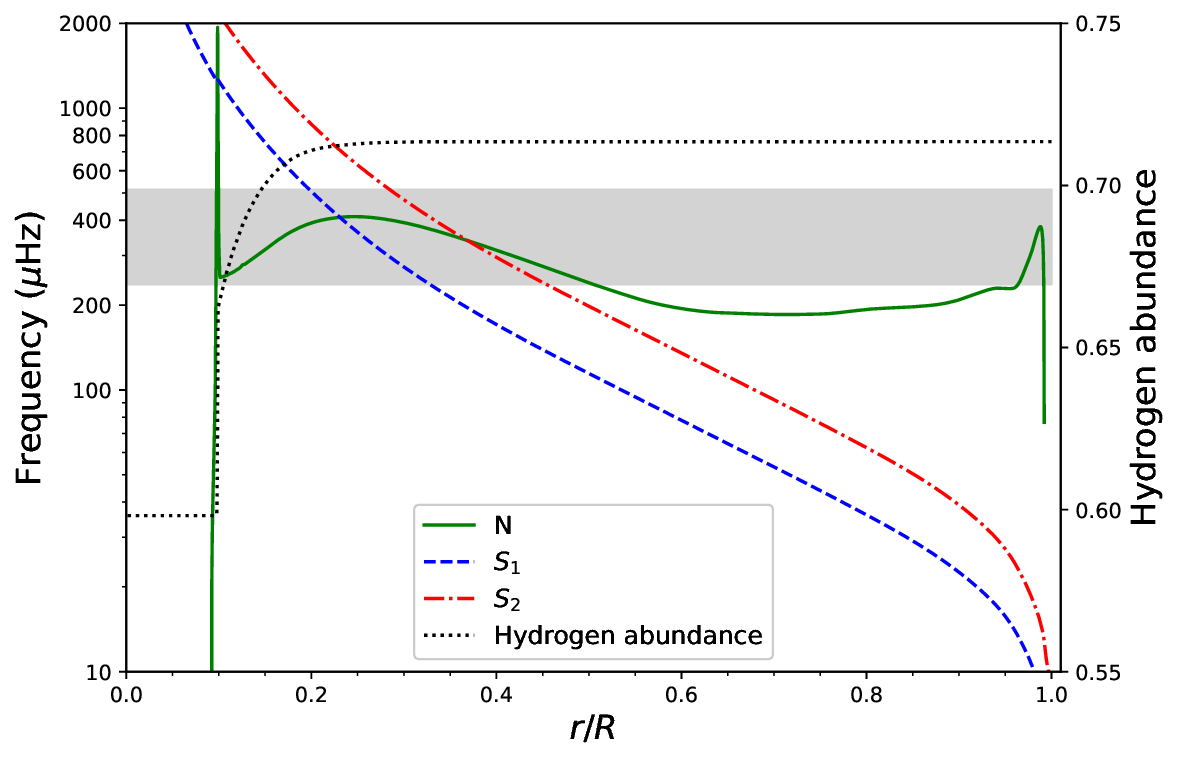}
  \caption{\label{Figure 9} Brunt$-$V$\ddot{\rm a}$is$\ddot{\rm a}$l$\ddot{\rm a}$ frequency $N$, the characteristic acoustic frequencies $S_{\ell}$ ($\ell$ = 1 and 2), and the hydrogen abundance inside the best-fitting model of TIC 120857354. The gray zone corresponds to the observed frequency range of TIC 120857354.}
\end{figure*}

\begin{figure*}
\centering
\includegraphics[width=\textwidth, angle = 0]{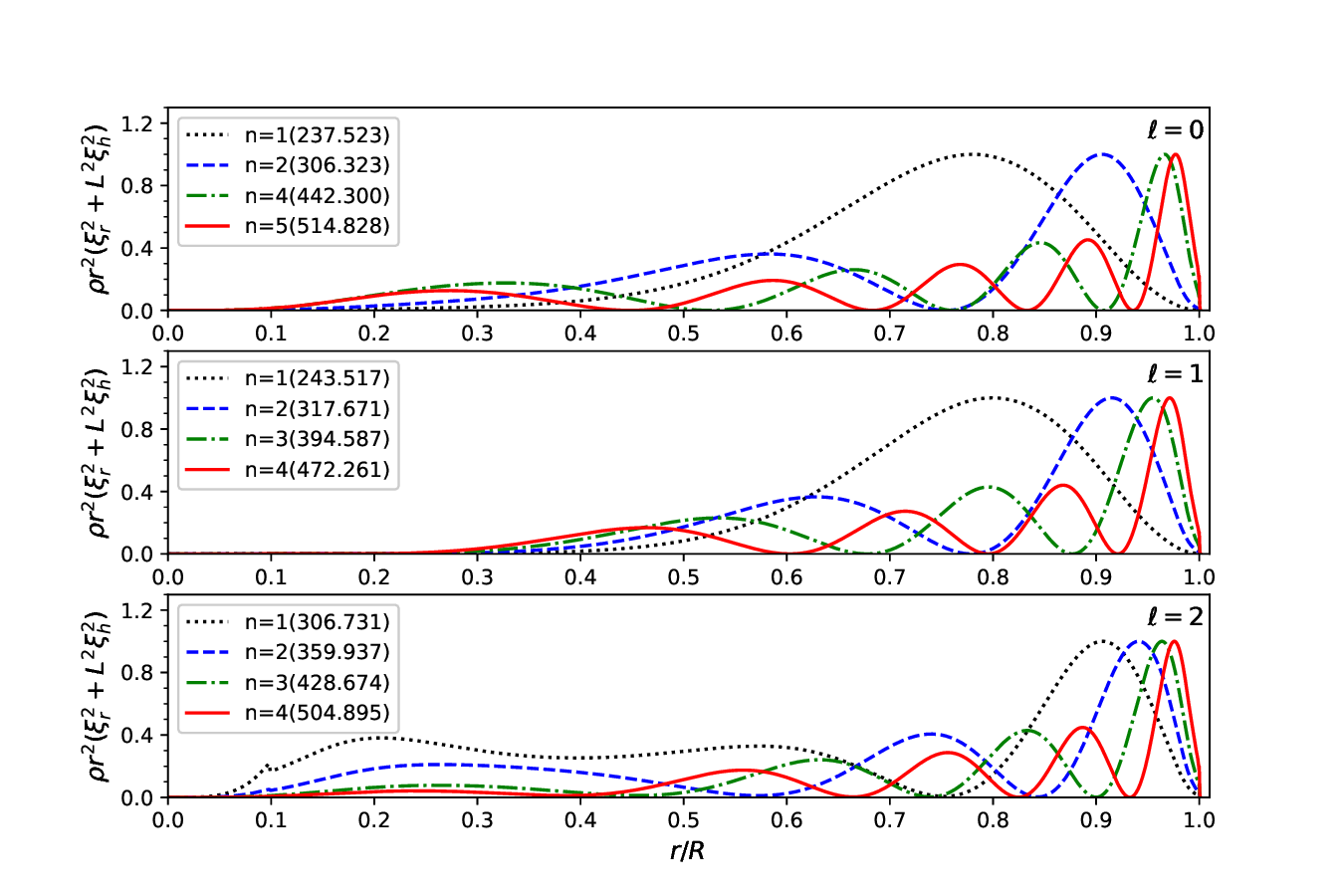}
  \caption{\label{Figure 10} Scaled kinetic energy weight functions inside the best-fitting model of TIC 120857354. }
\end{figure*}

\begin{figure*}
\centering
\includegraphics[width=\textwidth, angle = 0]{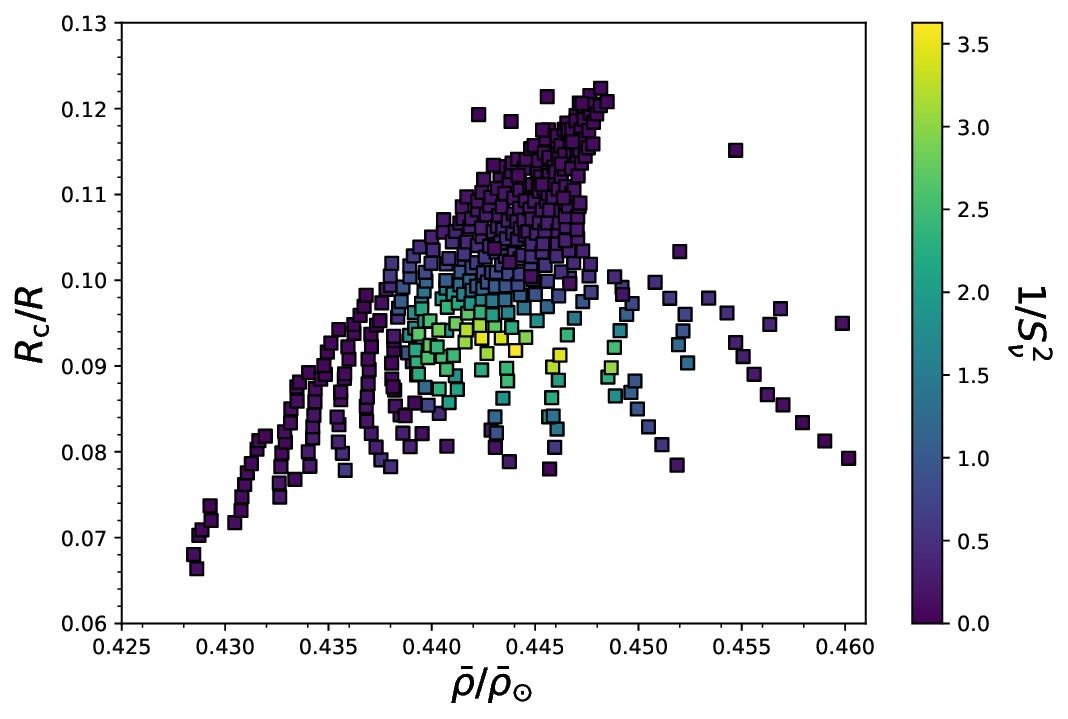}
  \caption{\label{Figure 11} Color diagram of the fitting results 1/S$_{\nu}^2$ versus the relative radius of the convective core $R_c$/$R$ and the mean density $\bar{\rho}/\bar{\rho}_{\odot}$. }
\end{figure*}

\begin{table*}
\centering
\footnotesize
\caption{\label{t2} Frequencies detected for TIC 120857354.}
\begin{tabular}{ccccccccccccccccc}
\hline\hline
 ID    &Freq.                 &Freq.             &Ampl.               &S/N   &Remark     &ID    &Freq.                 &Freq.             &Ampl.               &S/N    &Remark \\
        &(day$^{-1}$)   &($\mu$Hz)  &($\mu$mag)   &         &                 &        &(day$^{-1}$)   &($\mu$Hz)  &($\mu$mag)   &        & \\
\hline
$f_{ 1}$ &20.4924(  1) &237.180(  1) &38647( 25) & 79.10 &                 &$f_{27}$ &61.7236( 82) &714.394( 95) &  255( 22) & 10.80 &2$f_1+f_2$  \\
$f_{ 2}$ &20.7388(  2) &240.032(  3) & 9242( 22) &141.19 &                 &$f_{28}$ &38.2564( 92) &442.783(107) &  245( 23) & 10.29 &  \\
$f_{ 3}$ &26.4309(  9) &305.913( 10) & 6332( 44) & 20.75 &                 &$f_{29}$ &67.4145(104) &780.260(120) &  220( 22) &  8.60 &2$f_1+f_3$  \\
$f_{ 4}$ &26.4868( 10) &306.561( 12) & 4285( 50) & 31.17 &                 &$f_{30}$ &53.7853(108) &622.515(124) &  206( 22) &  7.48 &$f_4+f_7$  \\
$f_{ 5}$ &40.9847(  5) &474.361(  6) & 3973( 22) & 38.78 &2$f_1$           &$f_{31}$ &26.4623(286) &306.277(331) &  203( 26) &  5.44 &unresolved peak  \\
$f_{ 6}$ &41.2310( 13) &477.211( 15) & 1594( 23) & 46.56 &$f_1+f_2$        &$f_{32}$ &44.4227(120) &514.152(139) &  187( 23) &  7.13 &  \\
$f_{ 7}$ &27.2973( 14) &315.941( 16) & 1537( 21) & 16.25 &                 &$f_{33}$ &47.4462(128) &549.146(148) &  179( 21) &  5.64 &$f_4+f_{26}$  \\
$f_{ 8}$ &46.9238( 20) &543.100( 24) & 1230( 26) & 19.17 &$f_1+f_3$        &$f_{34}$ &31.5058(131) &364.650(151) &  177( 23) &  6.33 &$f_{15}-f_{12}$  \\
$f_{ 9}$ &26.4439( 60) &306.064( 69) &  922( 32) &  9.79 &unresolved peak  &$f_{35}$ &26.2626(132) &303.966(153) &  176( 24) &  5.38 &$f_{17}-f_2$  \\
$f_{10}$ &26.9222( 26) &311.600( 30) &  864( 22) & 12.06 &                 &$f_{36}$ &48.0024(138) &555.583(160) &  167( 22) &  6.02 &$f_1+f_{13}$  \\
$f_{11}$ &47.6620( 36) &551.644( 42) &  601( 22) &  9.71 &$f_2+f_{10}$     &$f_{37}$ &37.0294(158) &428.581(183) &  140( 21) &  5.98 &  \\
$f_{12}$ & 5.9389( 40) & 68.737( 47) &  581( 23) &  9.08 &$f_3-f_1$        &$f_{38}$ &43.7065(170) &505.863(196) &  139( 23) &  5.09 &  \\
$f_{13}$ &27.5076( 38) &318.375( 44) &  577( 23) & 14.84 &                 &$f_{39}$ &37.2331(162) &430.938(187) &  138( 22) &  5.31 &  \\
$f_{14}$ &26.4198(111) &305.785(129) &  563( 38) & 10.65 &unresolved peak  &$f_{40}$ &31.4256(187) &363.723(217) &  131( 24) &  5.13 &  \\
$f_{15}$ &37.4225( 41) &433.131( 47) &  519( 23) & 14.25 &                 &$f_{41}$ &43.2751(201) &500.869(233) &  129( 22) &  5.10 &  \\
$f_{16}$ &20.4795( 68) &237.031( 79) &  458( 26) &  8.79 &unresolved peak  &$f_{42}$ &47.5442(180) &550.280(208) &  128( 22) &  4.99 &$f_1+f_{24}$  \\
$f_{17}$ &46.9788( 57) &543.736( 66) &  456( 23) &  9.94 &$f_1+f_{4}$      &$f_{43}$ &34.1101(206) &394.793(238) &  118( 21) &  4.49 &  \\
$f_{18}$ &47.7889( 51) &553.112( 59) &  429( 22) & 10.47 &$f_1+f_7$        &$f_{44}$ &46.9358(261) &543.238(302) &  116( 26) &  4.53 &$f_1+f_3$   \\
$f_{19}$ &61.4774( 51) &711.544( 59) &  414( 21) & 13.33 &3$f_1$           &$f_{45}$ &41.4775(216) &480.064(250) &  109( 21) &  4.14 &2$f_2$  \\
$f_{20}$ &36.8176( 56) &426.130( 65) &  397( 22) & 12.75 &                 &$f_{46}$ &33.9047(236) &392.415(273) &  107( 22) &  4.41 &  \\
$f_{21}$ &40.5608( 61) &469.454( 70) &  350( 23) & 12.02 &                 &$f_{47}$ &36.5954(254) &423.558(294) &  104( 22) &  4.49 &  \\
$f_{22}$ &20.5079( 99) &237.360(115) &  322( 24) &  6.91 &unresolved peak  &$f_{48}$ &61.0497(240) &706.594(278) &  104( 21) &  4.62 &$f_1+f_{21}$  \\
$f_{23}$ &47.1723( 73) &545.976( 85) &  302( 22) &  8.75 &$f_2+f_3$        &$f_{49}$ &67.6606(237) &783.109(274) &   98( 22) &  4.44 &$f_1+f_{23}$   \\
$f_{24}$ &27.0535( 75) &313.119( 87) &  301( 23) &  7.79 &$f_{18}-f_2$     &$f_{50}$ &68.1564(275) &788.848(318) &   97( 23) &  4.16 &$f_1+f_{11}$   \\
$f_{25}$ &26.4928(175) &306.630(203) &  301( 53) &  7.35 &unresolved peak  &$f_{51}$ &67.4719(340) &780.924(394) &   89( 23) &  4.37 &2$f_1+f_4$  \\
$f_{26}$ &20.9760( 79) &242.778( 91) &  289( 23) &  7.59 &2$f_2-f_1$       &  \\
\hline
\end{tabular}
\end{table*}
\newpage

\begin{table*}
\caption{\label{t2}Potential rotational splittings. $\delta\nu$ denotes splitting frequency in $\mu$Hz.}
\centering
\begin{tabular}{llllllll}

\hline\hline

Group    &ID        &Freq.          &$\delta\nu$ &$l$ &$m$ \\

          &          &($\mu$Hz)      &($\mu$Hz)   &    &\\
\hline
          &$f_{47}$  &423.558        &            &2    &$-2$ \\
          &          &               &2.572\\
          &$f_{20}$  &426.130        &            &2    &$-1$ \\
          &          &               &2.451\\
  1       &$f_{37}$  &428.581        &            &2    &$0$ \\
          &          &               &2.357\\
          &$f_{39}$  &430.938        &            &2    &$+1$ \\
          &          &               &2.193\\
          &$f_{15}$  &433.131        &            &2    &$+2$ \\
          &\\ 
          &\\
          &$f_{4}$   &306.561        &            &1 or 2\\
  2       &          &               &5.039       &\\
          &$f_{10}$  &311.600        &            &1 or 2\\
          &\\ 
          &\\ 
          &$f_{7}$   &315.941        &            &1 or 2\\
  3       &          &               &2.434       \\
          &$f_{13}$  &318.375        &            &1 or 2\\
          &\\
          &\\
          &$f_{46}$  &392.415        &            &1 or 2\\
  4       &          &               &2.378       \\
          &$f_{43}$  &394.793        &            &1 or 2\\
          &\\
          &\\
          &$f_{41}$  &500.869        &            &1 or 2\\
  5       &          &               &4.994\\
          &$f_{38}$  &505.863        &            &1 or 2\\
\hline

\end{tabular}

\end{table*}
\begin{table*}
\caption{\label{t2} Results of mode identifications. $\Delta\nu$ represents frequency spacing in $\mu$Hz}
\centering
\begin{tabular}{llllllll}
\hline\hline
  ID        &Freq.      &$\Delta\nu$  &$l$ &$n$ &$m$ \\
            &($\mu$Hz)  &($\mu$Hz)    &    &    &\\
\hline
$f_{1}$    &237.180     &            &0    &1 & \\
            &           &68.733\\
$f_{3}$    &305.913     &            &0    &2 &\\
            &           &136.817\\
$f_{28}$   &442.783     &            &0    &4 &\\
            &           &71.369\\
$f_{32}$   &514.152     &            &0    &5 &\\
\hline
$f_{2}$    &240.032     &           &1    &&?\\
           &            &75.909     \\
$f_{7}$    &315.941     &           &1    &&($-1$, 0)\\
$f_{13}$   &318.375     &           &1    &&(0, $+1$)\\
            &           &74.040  \\
$f_{46}$   &392.415     &          &1    &&($-1$, 0)\\
$f_{43}$   &394.793     &          &1    &&(0, $+1$)\\
            &           &74.661  \\
$f_{21}$   &469.454     &          &1    &&?\\
\hline
$f_{4}$   &306.561      &           &2    &&($-2$, $-1$, 0) \\
$f_{10}$  &311.600      &           &2    &&(0, $+1$, $+2$) \\
          &             &52.123\\
$f_{40}$  &363.723      &           &2    &&?\\
          &             &64.858\\
$f_{47}$  &423.558      &           &2    &&$-2$ \\
$f_{20}$  &426.130      &           &2    &&$-1$ \\
$f_{37}$  &428.581      &           &2    &&$0$ \\
$f_{39}$  &430.938      &           &2    &&$+1$ \\
$f_{15}$  &433.131      &           &2    &&$+2$ \\
            &           &72.288\\
$f_{41}$  &500.869      &           &2    &&($-2$, $-1$, 0)\\
$f_{38}$  &505.863      &           &2    &&(0, $+1$, $+2$)\\

\hline

\end{tabular}

\end{table*}

\begin{table*}
\centering
\caption{\label{t3}Fundamental parameters derived from asteroseismology.}
\begin{tabular}{lccccc}
\hline\hline
Parameters                      &Values             \\
\hline
$M$ ($M_{\odot}$)               &1.54 $\pm$ 0.04         \\
$Z$                                     &0.015 $\pm$ 0.003       \\
$T_{\rm eff}$ (K)               &7441 $\pm$ 370          \\
$\log g$                            &4.27 $\pm $ 0.01        \\
$R$ ($R_{\odot}$)               &1.52 $\pm$ 0.01         \\
$L$ ($L_{\odot})$               &6.33 $\pm$ 1.53         \\
Age (Gyr)                       &0.53 $\pm$ 0.07         \\
$X_{\rm c}/X_0$                     &0.84 $\pm$ 0.05          \\
$\bar{\rho}/\bar{\rho}_{\odot}$ &0.444 $\pm$ 0.004        \\
$R_{\rm conv}/R$                &0.092 $\pm$ 0.002        \\
\hline
\end{tabular}
\end{table*}

\begin{table*}
\centering
\caption{\label{t4}Theoretical frequencies of the best-fitting model. $\nu_{\rm mod}$ is the model frequency. $\ell$ and $n$ are its spherical harmonic degree and radial order, respectively. $\beta_{\ell,n}$ is the rotational parameters, whose values are determined by distribution of the radial displacement and the horizontal displacement in the star. $E_g$  represents mode inertia in the g-mode propagation region, and $E$ is the total mode inertia. }
\label{observed frequencies}
\begin{tabular}{ccccccccc}
\hline\hline
$\nu_{\rm mod}(\ell,n)$ &$E_{\rm g}/E$ &$\nu_{\rm mod}(\ell,n)$ &$\beta_{\ell, n}$ &$E_{\rm g}/E$ &$\nu_{\rm mod}(\ell,n)$ &$\beta_{\ell, n}$ &$E_{\rm g}/E$\\
 ($\mu$Hz)              &                 &($\mu$Hz)               &                  &                 &($\mu$Hz)               &&\\
\hline
237.523(0, 1) &0     &169.546(1,-1) &0.528 &0.561   &193.525(2,-2) &0.807 &0.724\\
306.323(0, 2) &0     &243.517(1, 1) &0.989 &0.011   &229.847(2,-1) &0.876 &0.451\\
374.113(0, 3) &0     &317.671(1, 2) &0.995 &0.007   &257.237(2, 0) &0.958 &0.342\\
442.300(0, 4) &0     &394.587(1, 3) &0.993 &0.003   &306.731(2, 1) &0.921 &0.267\\
514.828(0, 5) &0     &472.261(1, 4) &0.990 &0       &359.937(2, 2) &0.841 &0.185\\
591.388(0, 6) &0     &550.663(1, 5) &0.987 &0       &428.674(2, 3) &0.904 &0\\
668.851(0, 7) &0     &629.151(1, 6) &0.985 &0       &504.895(2, 4) &0.941 &0\\
746.684(0, 8) &0     &707.048(1, 7) &0.984 &0       &583.646(2, 5) &0.959 &0\\
825.607(0, 9) &0     &785.102(1, 8) &0.984 &0       &662.105(2, 6) &0.969 &0\\
906.244(0,10) &0     &864.498(1, 9) &0.984 &0       &740.233(2, 7) &0.974 &0\\
988.410(0,11) &0     &945.517(1,10) &0.984 &0       &819.017(2, 8) &0.978 &0\\

\hline
\end{tabular}
\end{table*}

\begin{table*}
\caption{\label{t2} Results of comparisons of frequencies for those modes in Table 3. $\nu_{\rm obs}$ denotes observed frequency and $\nu_{\rm mod}$ denotes model frequency.}
\centering
\begin{tabular}{lccccccc}
\hline\hline
  ID        &$\nu_{\rm obs}$   &$\nu_{\rm mod}$($l$, $n$, $m$) &$\nu_{\rm obs}$-$\nu_{\rm mod}$\\
            &($\mu$Hz)         &($\mu$Hz)                      &($\mu$Hz)     \\
\hline
$f_1$          &237.180 &237.523(0,1, 0) &-0.343 &\\
$f_3$          &305.913 &306.323(0,2, 0) &-0.411 &\\
$f_{28}$      &442.783 &442.300(0,4, 0) &0.482 &\\
$f_{32}$      &514.152 &514.828(0,5, 0) &-0.676 &\\
\hline
$f_2$          &240.032 &241.102(1,1,-1) &-1.069 &\\
$f_7$          &315.941 &315.241(1,2,-1) &0.701 &\\
$f_{13}$     &318.375 &317.671(1,2, 0) &0.704 &\\
$f_{46}$     &392.415 &392.161(1,3,-1) &0.254 &\\
$f_{43}$     &394.793 &394.586(1,3, 0) &0.206 &\\
$f_{21}$     &469.454 &469.843(1,4,-1) &-0.390 &\\
\hline
$f_{4}$       &306.561 &306.731(2,1, 0) &-0.170 &\\
$f_{10}$     &311.600 &311.229(2,1, 2) &0.371 &\\
$f_{40}$    &363.723 &364.045(2,2, 2) &-0.322 &\\
$f_{47}$    &423.558 &424.259(2,3,-2) &-0.701 &\\
$f_{20}$    &426.130 &426.466(2,3,-1) &-0.336 &\\
$f_{37}$    &428.581 &428.674(2,3, 0) &-0.093 &\\
$f_{39}$    &430.938 &430.882(2,3, 1) &0.057 &\\
$f_{15}$    &433.131 &433.089(2,3, 2) &0.041 &\\
$f_{41}$    &500.869 &500.298(2,4,-2) &0.570 &\\
$f_{38}$    &505.863 &504.895(2,4, 0) &0.968 &\\
\hline

\end{tabular}

\end{table*}

\label{lastpage}
\end{document}